\documentclass[times,twocolumn]{aastex63}

\usepackage{url}

\received{April 7, 2020}
\revised{October 6, 2020}
\accepted{October 6, 2020}
\published{November 20, 2020}

\shorttitle{Top-Heavy IMF}
\shortauthors{Haghi {\em et al.}}

\begin{document}
	
	\title{The Lifetimes of Star Clusters Born with a Top-heavy IMF}
	
	\correspondingauthor{Hosein Haghi}
	\email{haghi@iasbs.ac.ir}
	
	\author[0000-0002-9058-9677]{H.~Haghi}
	\affiliation{Department of
		Physics, Institute for Advanced Studies in Basic Sciences (IASBS), Zanjan 45137-66731, Iran}
	
	\author[0000-0001-6449-8758]{G.~Safaei}
	\affil{Department of
		Physics, Institute for Advanced Studies in Basic Sciences (IASBS), Zanjan 45137-66731, Iran} 
	
	\author[0000-0002-0322-9957]{A.~H.~Zonoozi}
	\affiliation{Department of
		Physics, Institute for Advanced Studies in Basic Sciences (IASBS), Zanjan 45137-66731, Iran}
	
	\author[0000-0002-7301-3377]{P.~Kroupa}
	%\affiliation{Argelander Institut f\"ur Astronomie(AIfA), Auf dem H\"ugel 71, Bonn, D-53121, Germany}
	\affiliation{Helmholtz-Institut f\"ur Strahlen-und Kernphysik (HISKP), Universit\"at Bonn, Rheienische Friedrich-Wilhelms Universit\"at Nussallee 14-16, Bonn, D-53115, Germany}
	\affiliation{Charles University in Prague, Faculty of Mathematics and Physics, Astronomical Institute, V Holesovickach 2, CZ-180 00 Praha 8,  Czech Republic}

	\begin{abstract}
		
		Several observational and theoretical indications suggest that the initial mass function (IMF) becomes increasingly top-heavy (i.e., overabundant in high-mass stars with mass $m > 1M_{\odot}$) with decreasing metallicity and increasing gas density of the forming object. This affects the evolution of globular clusters (GCs) owing to the different mass-loss rates and the number of black holes  formed. Previous numerical modeling of GCs usually assumed an invariant canonical  IMF. Using the state-of-the-art \textsc{nbody6} code, we perform a comprehensive series of direct $N$-body simulations to study the evolution of star clusters, starting with a top-heavy IMF and undergoing early gas expulsion. Utilizing the embedded cluster mass-radius relation of Marks \& Kroupa (2012) for initializing the models, and by varying the degree of top-heaviness, we calculate the minimum cluster mass needed for the cluster to survive longer than 12 Gyr.  We document how the evolution of different characteristics of star clusters such as the total mass,  the final size, the density, the mass-to-light ratio, the population of stellar remnants, and the  survival  of GCs is influenced by  the degree of top-heaviness. We find that the lifetimes of clusters with different IMFs  moving on the same orbit are proportional to the relaxation time to a power of $x$ that is in the range of 0.8 to 1. The observed correlation between concentration and the mass function slope in Galactic GCs can be accounted for excellently in models starting with a top-heavy IMF and undergoing an early phase of rapid gas expulsion.
		
		\end{abstract}

	\section{Introduction} \label{sec:intro}
	
	Globular clusters (GCs) lose stars as a result of both internal dynamics and  the presence of the external tidal field from the host galaxy. The mass loss is due to stellar and binary evolution, evaporation driven by two-body relaxation, ejection of stars caused by few-body interactions, and tidal stripping imposed by the gravitational field of the host galaxy. While the external effects might ultimately destroy GCs, the internal effects such as stellar evolution, stellar mass segregation, stellar remnants, and  binaries can remove stars from GCs, altering their final dynamical structure and providing escaped stars that contribute to the host galaxy. Disentangling and constraining processes that determine the survival and destruction rate  of a GC is one of the most challenging issues. 
	
	The fate of a star cluster and the stars it loses to their  host galaxy also depends on its initial condition. For example, a tidally filling cluster that is initially set up primordially mass segregated will expand twice more (caused by the early evolution of massive stars) compared to a similar cluster but  without primordial mass segregation,  and consequently will dissolve faster \citep{ BaumgardtDK2008, Vesperini_2009, Haghi14}. Tidally underfilling clusters, however, can survive this early expansion
	and have a lifetime similar to that of unsegregated clusters. The origin and effect of primordial mass segregation have been widely discussed on the basis of both observational \citep{ Brandl_1996, Hillenbrand_1998, deGrijs_2002,  Frank_2012} and theoretical grounds \citep{Bonnell_1997, McMillan_2007, Haghi14, Haghi15, Pavlik19}.

	Moreover, in young, still gas-embedded star clusters, the rapid expulsion of the residual gas that did not form stars causes star clusters to become supervirial on a dynamical timescale, and hence a fraction of the originally bound stars become unbound and leave the cluster.  %The gas is lost over a time which is determined by feedback from massive stars through UV radiation and massive stellar winds from OB stars or supernova explosions.
	The gas expulsion strongly affects further dynamical evolution of star clusters. Studying the dynamical effect of gas dispersal by adopting a time-varying background potential mimicking the influence of the gas is commonly used in such studies because numerically treating a hydrodynamical component with radiation transfer is computationally not reachable. Using a large set of $N$-body integrations, \cite{BaumgardtKroupa2007} showed that the survival rate and final properties of star clusters are strongly influenced by the parameter values for the gas expulsion timescale, $\tau_{GE}$, the star formation efficiency (SFE), $\varepsilon$, and the ratio of the initial half-mass radius to tidal radius.  By choosing the appropriate values for these parameters, \cite{Banerjee2013} succeeded in explaining well-observed, very young massive clusters such as R136 and  NGC 3603. Noteworthy here is that the same values for $\varepsilon \simeq \frac{1}{3}$ and $\tau_{GE}\simeq$ thermal timescales are needed for the Orion Nebula Cluster \citep{KroupaArsethHurley2001,Kroupa_2018}. This indicates a certain universality of the formation process of star clusters over the mass range  $10^3-10^5 M_{\odot}$. The impact of the early gas expulsion and strong mass loss by stellar evolution on the early dynamical evolution of initially massive compact clusters and their survivability  has been investigated by \cite{Brinkmann2017}. The flattening of the low-mass  stellar mass function (MF) slope driven by a violent early phase of gas expulsion from an embedded cluster with primordial mass segregation can also be a possible explanation for the observed flattened MF of some remote halo GCs, such as Palomar 4 and Palomar 14 \citep{Zonoozi11, Zonoozi14, Zonoozi17, Haghi15}. 
	
	The initial MF (IMF) of stars within an embedded cluster is also one of the most important initial conditions that plays a significant role in star cluster evolution. Many observational and theoretical efforts have been made to constrain the shape of the stellar IMF
	\citep{Salpeter1955, MillerScalo1978, Scalo1986, Kroupa2001, Chabrier2003,  Bastian2010, DeMarchi_2010, Hopkins2012, Kroupa2013, Offner_2014}. Most studies of resolved stellar populations in the disk of the Milky Way showed that stars form following an IMF that has a universal form  \citep{Kroupa2001, Kroupa2002, Bastian2010} which is referred to as the "canonical" IMF. This poses a problem for star formation theories that predict a dependence on the environment where star formation takes place \citep{Kroupa2013}. 
	
	The shape of the stellar IMF of a star cluster near its upper mass limit is a focal topic of investigation, as it determines the high-mass stellar content and hence the luminosity and dynamics of the cluster in its embedded phase.  While many aspects of the residual gas expulsion in  the embedded cluster with the universal IMF have already been studied in the literature, a systematic study of the survival rate of the star clusters evolving under residual gas expulsion with a systematically varying IMF is not yet available. In this paper, we aim at considering the pure effect of the top-heaviness of the IMF on the early and long-term evolution of GCs undergoing gas expulsion. We investigate the influence of the top-heaviness of the IMF on the dissolution/survival rate of GCs, also addressing the question of the contribution of clustered star formation to the population of field stars in a galaxy.  We calculate a grid of models over a  wide range of stellar masses, undergoing primordial residual gas expulsion, and with the IMF slope varying in the high-mass range ($\alpha_3$).

	The paper is organized as follows:  In Section \ref{sec:Initial Conditions}, we describe the initial setup of the $N$-body models and our simulation method. The main results for the dissolution rate of star clusters and the evolution of cluster parameters  are presented in Section \ref{sec:Results}.  Finally, Section \ref{sec:conclusions}  consists of a discussion and the conclusions.
	
	\section{Observational evidence for a systematically varying IMF}
	
	Suggestions for the IMF not being universal have accumulated for different types of stellar systems such as galaxies (e.g. \citealt{Matteucci_1994, Baugh2005, Nagashima2005, vanDokkum2008, Hoversten08, Lee09,   Meurer09, Gunawardhana11}) and the Galactic bulge and center (e.g. \cite{Ballero2007}; \cite{Maness2007}). Several observational and theoretical indications suggest that the IMF becomes top-heavy (i.e., overabundant in high-mass stars)  under extreme starburst conditions
	\citep{Dabringhausen2009, Dabringhausen2010, Dabringhausen2012, MarksKroupaDabringhausen2012, Kroupa2013, Zhang18}. The data suggest that the IMF becomes more top-heavy with decreasing cluster metallicity and increasing density  (i.e., the high-mass IMF slope is flatter in denser and metal-poorer environments).
	
	\cite{Adams1996} developed a model of self-regulated accretion onto protostars implying higher accretion rates at lower metallicities and thus larger stellar masses.  \cite{Lin1996} discussed interactions of prestellar clumps leading to mergers as a process in star formation. Their model predicts an increase in the mean stellar mass with the density of the star-forming region.  Moreover, in very dense star-forming cores, prestellar cores may coalesce before they form protostars, thus leading to a top-heavy IMF \citep{Dib2007a, Dib2007b}.  On the other hand, in a gas cloud of low metallicity, the Jeans mass is larger, favoring the formation of more massive stars, and the fraction of high-mass to low-mass stars increases \citep{Larson1998}.

	The need of IMF variation has been also put forward to interpret the evidence coming from mass-to-light ratios estimated through integrated light analysis of M31 GCs \citep{Haghi_2017}. By taking into account dynamical evolution, together with a top-heavy IMF and a 10\% retention fraction for remnants, a better agreement to the data can be obtained \citep{Zonoozi16}, and  incorporating the age-metallicity relation further improves the fit \citep{Haghi_2017}. 
	
	Moreover, \cite{Banerjee2012} found that the  top-heaviness of the true high-mass IMF over the observationally determined one is a general feature of massive, young clusters, where the dynamical ejection of massive stars is efficient \citep{Oh_2015, Oh_2016}. They showed that if the measured IMF of the R136 young massive cluster is granted to be canonical, as observations indicate, then the true high-mass IMF of R136 at its birth must be at least moderately top-heavy when corrected for the dynamical escape of massive stars. A top-heavy IMF in R136 is further supported by a direct star-counts analysis \citep{Bestenlehner2020}.

	In addition, the reported range of the MF slopes of metal-rich young clusters is very close to the canonical value of 2.3, which is consistent with those expected from empirical top-heavy IMF formulae (Equation (15) of \citealt{MarksKroupaDabringhausen2012}).  For example, the reported MF slope for the NGC 3603 cluster is $\alpha_3=$1.9 - 2.3 \citep{Nurenberger_2002, Sung_2004, Stolte_2006, Harayama_2008}. R136, which has been mentioned above,  the whole 30 Dor star-forming region in the Large Magellanic Cloud \citep{Schneider18}, and the metal-poor very young massive cluster NGC 796 \citep{Kalari18} have been suggested to have a top-heavy IMF. Moreover, the Arches cluster near the Galactic center has been formed to have  canonical IMF \citep{Park_2020} in agreement with  \citet{MarksKroupaDabringhausen2012}.

	%R136 has  $\alpha_3=$ 2.2 to 2.6 \citep{Brandl_1996, Massey_1998, Andersen_2009}, but is likely to have been top-heavy \citep{Banerjee2012, Bestenlehner2020}, and the 30 Dor region in the Large Magellanic Cloud also exhibits no deviation from the canonical IMF \citep{Selman_2005}. It should be pointed out that our results in this paper are based on the  assumption of the top-heavy IMF that are questioned however by some authors for star clusters evolving within a strong tidal filed of Galactic center and argued that it is possible to dynamically form star clusters in the form of the top-heavy MF from a region with a canonical IMF.  Using $N$-body simulation, \citealt{Park_2020}  showed that many low mass stars are lost in a strong tidal field of Galactic center, leading to the surviving clusters to contain an overabundance of massive stars and appear to have top-heavy IMFs. 

	%Direct star count in the 30 Dor star forming region and in a young cluster in the Small Megellanic Cloud also indicate a top-heavy IMF \citep{Schneider18, Kalari18}. 

	%%%%%%%
	\begin{figure}[]
		\centering
		\includegraphics[width=8cm]{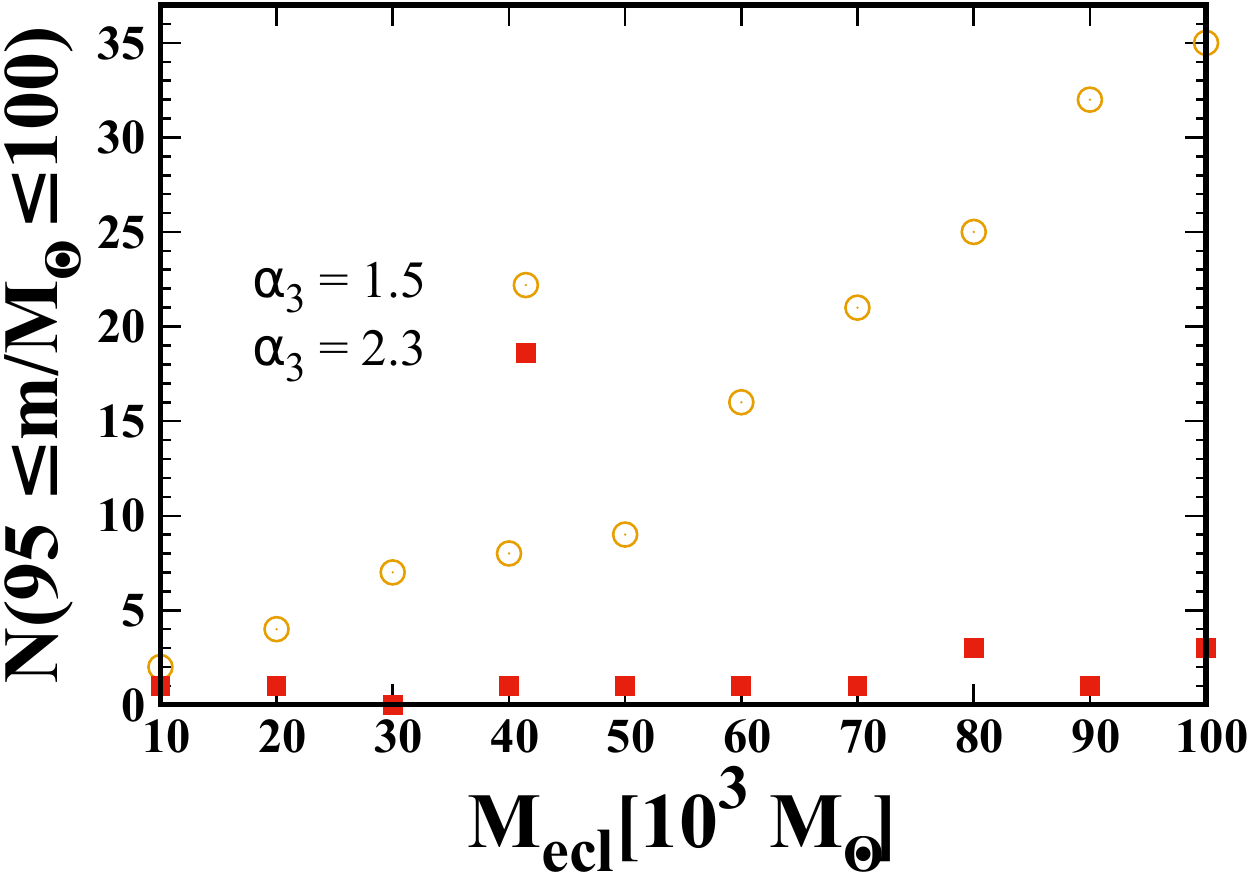}
		\includegraphics[width=8cm]{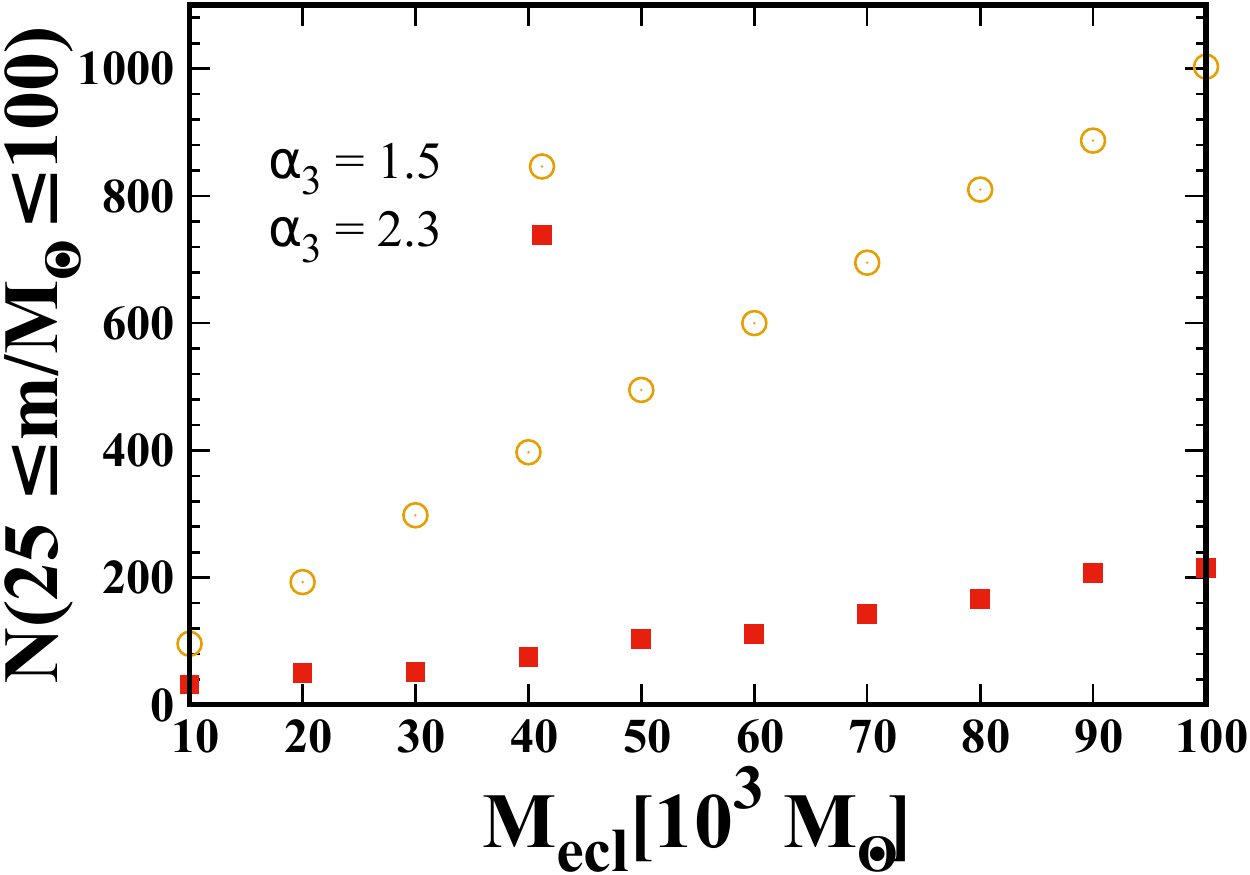}
		\caption{Initial number of massive stars in the range from 95 to 100$M_{\odot}$ (top panel) and from 25 to 100$M_{\odot}$ (bottom panel) for models with different initial masses and different values of the $\alpha_3$ parameter.  Although the value of maximum stellar mass in each model is different, the mean value of the generated maximum stellar mass for all models is 99.4$M_{\odot}$. }
		\label{fig:mstar-max-25-95-100}
		\end{figure}
	
	\begin{figure*}[]
		\centering
		\includegraphics[width=17cm]{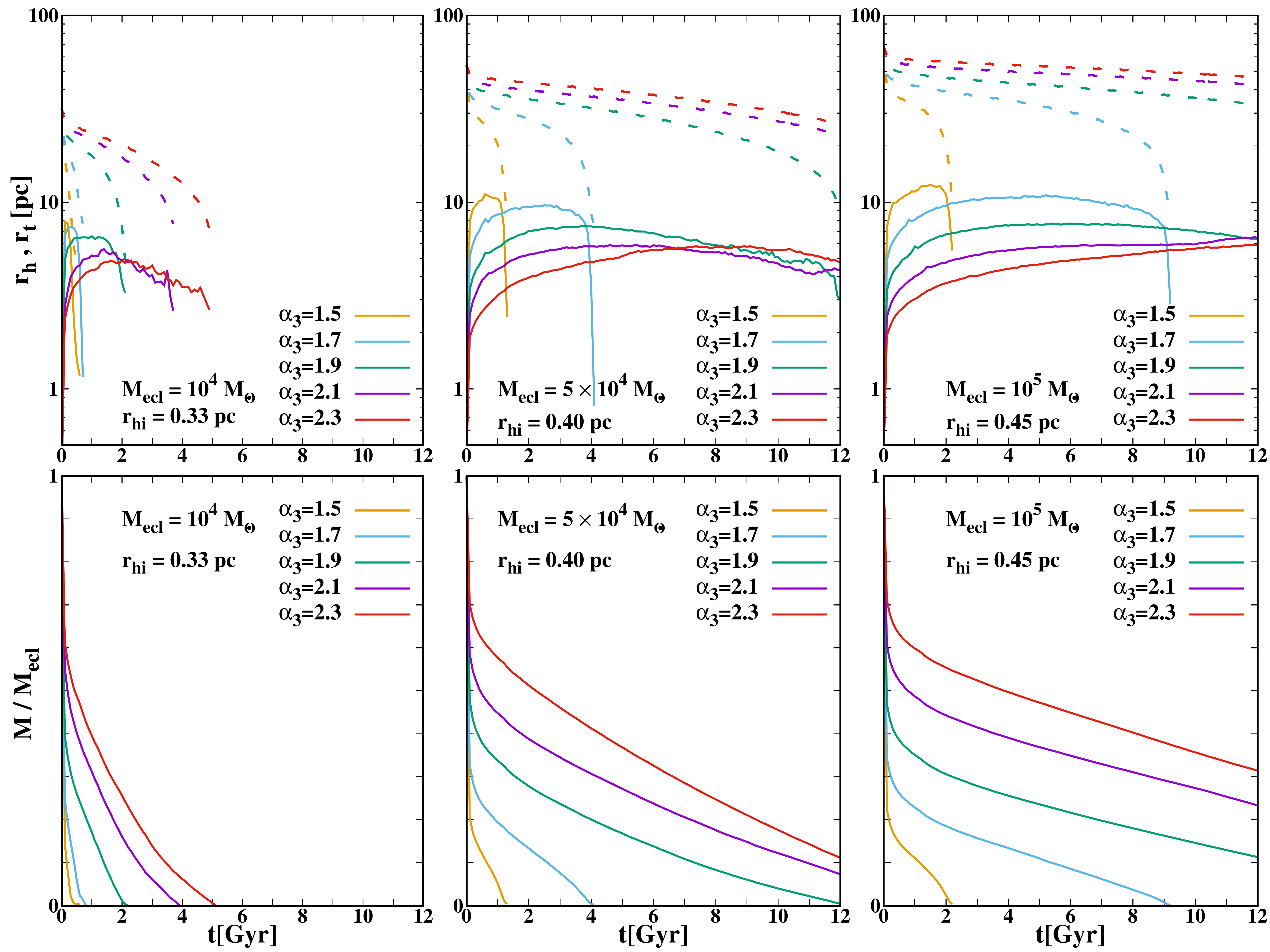}
		\caption{Top panels: evolution of the 3D half-mass radius (solid lines) and tidal radius (dashed lines) of clusters  undergoing residual gas expulsion for different initial values for $\alpha_3$ for three representative cases. Because of the stellar evolution, the clusters experience a more rapid expansion when the IMF is top-heavy than when it is canonical (red curves).  Bottom  panels:  total mass of the clusters as a function of time.  }
		\label{fig:m-rh-time-fig1}
		\end{figure*}
	
	\begin{figure*}[t]
		\centering
		\includegraphics[width=13cm]{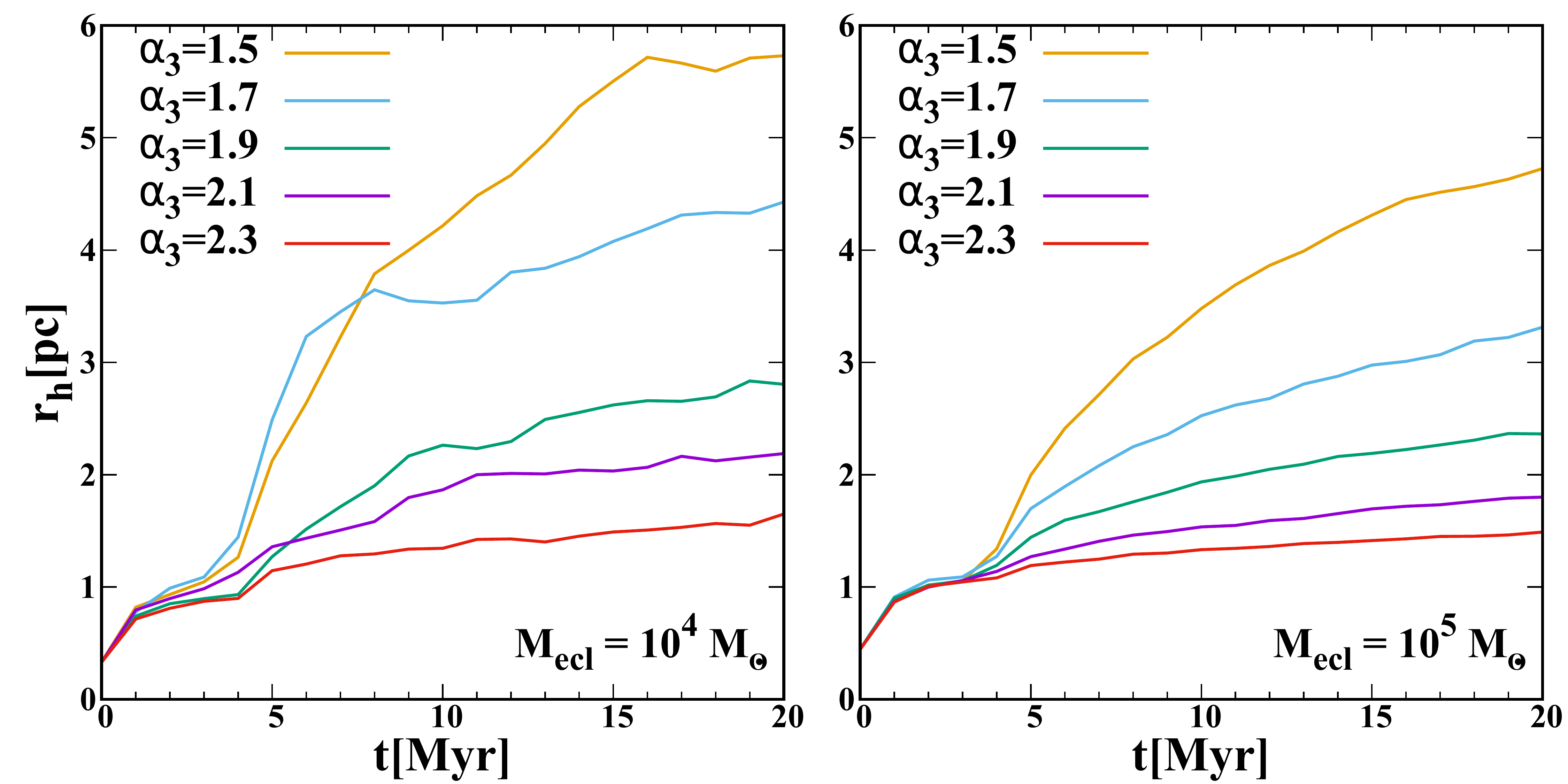}
		\caption{Time evolution of the half-mass radii for two computed $N$-body models with different initial masses and half-mass radii. From bottom to top, the curves represent different values for $\alpha_3$.    At a cluster age around 4 Myr, the first supernova explosion occurs, resulting in a more pronounced mass loss and an accelerated  increase of the half-mass radius.}\label{fig:rh-horizontal-5Myr-fig2}
		\end{figure*}
	
	\section{DESCRIPTION OF THE models and initial conditions} \label{sec:Initial Conditions}
	
	We use the state-of-the-art freely available collisional  direct $N$-body code  \textsc{nbody6}\footnote{\url{https://people.ast.cam.ac.uk/~sverre/web/pages/nbody.htm
	}} \citep{Aarseth2003, Nitadori2012} to evolve model star clusters from zero age to 12 Gyr, over a range of  initial half-mass radii, embedded cluster masses, and stellar IMF slopes in the high-mass range (i.e., $m> 1M_{\odot}$).  The dissolution time, $T_{diss}$, is defined to be the time when 95\% of the initial number of stars are lost from the cluster. For the initial density profile of the clusters we consider the Plummer profile \citep{Plummer1911} in virial equilibrium. Initially, dense spherical Plummer conditions are a good approximation given the structure of well-observed very young clusters \citep{KroupaArsethHurley2001, Banerjee2012, Banerjee2012a}.  The models are not initially mass segregated. All clusters move on circular orbits with circular velocity $V_G=220$ km~s$^{-1}$ at Galactocentric radius $R_G = 8.5$ kpc through the host galaxy that follows a phantom dark matter Milky Way-like potential (modeled as an NFW potential with $c=5$ as in  \cite{Kupper2015}).  
	
	We performed simulations with different initial cluster masses in the range $M_{ecl}=10^4$ to $3\times 10^5 M_{\odot}$. The corresponding size scale of the Plummer models was set by the initial $3D$ half-mass radius, $r_{hi}$, following \cite{MarksKroupa2012}:
	\begin{eqnarray}
		r_{hi}=0.1 \times \left(\frac{M_{ecl}}{M_{\odot}}\right)^{0.13} \rm pc.
		\label{equ:marks-kroupa-relation}
		\end{eqnarray}
	Therefore, the initial 3D half-mass radii of all models vary from 0.33 to 0.52 pc.
	
	The code  also includes a comprehensive treatment of single and binary stellar evolution from the zero-age main sequence (ZAMS) through remnant phases by using the \textsc{SSE/BSE} routines and analytical fitting functions developed by \cite{Hurley2000, Hurley2002}. The range of stellar masses was chosen to be from 0.08 to 100$M_{\odot}$ and the metallicity of the clusters is Z$ = 0.001$.  We computed one more model with an upper mass of 150\,$M_{\odot}$ to assess the effect of the maximum stellar mass on the evolution of star clusters. All stars are assumed to be on the ZAMS when the simulation begins at $t=0$.

	We have no primordial binaries in our models. However, all types of binaries, such as binary black holes (BHs), binary neutron stars (NSs), binary white dwarfs (WDs), and higher multiplicity systems, are allowed to form during the calculations. Some of these dynamically formed binaries are retained in the clusters over their entire evolution. 
	
	The starting point for all models is a star cluster in its gas-embedded phase. The residual gas within a newly formed star cluster is expelled through stellar feedback on a timescale $\leq$ 1 Myr. Gas expulsion is assumed to start at a certain time $t_D$, which is set equal to 0.6 Myr, which accounts for the ultracompact HII phase \citep{KroupaArsethHurley2001, Banerjee2013}. The SFE, i. e., the fraction of the gas that is converted into the stars in dense cores, can be defined as follows:
	\begin{equation}
		\epsilon = \frac{M_{ecl}}{M_{ecl}+M_{gas}},
		\label{equ:efficiency}
		\end{equation}
	where $M_{ecl}$ is the total mass of the stars formed within the embedded clusters and $M_{gas}$ is the mass of the gas not converted into stars \citep{BaumgardtKroupa2007}. The SFE is assumed to be equal to $\epsilon = 0.33$ in all models, which is consistent with the observationally determined value (\cite{Lada2003}, \cite{Megeath2016}). We assume that the SFE does not depend on the position inside the cluster, so gas and stars follow the same density distribution initially, which is given by a Plummer model.
	
	The gas is not simulated directly; instead, its influence on the stars is modeled as a modification to the equation of motion of the stars.
	%The gas expulsion starts after the embedded phase at a certain time, $t_D$ (delay time), which is set equal to 0.6 Myr.
	The mass of gas decreases exponentially with  a characteristic timescale of  $t_{decay}$ as follows:
	\begin{equation}
		M_{gas}(t)=M_{gas}(0) e^{-(t-t_D)/t_{decay}},  \label{equ:gas-mass}
		\end{equation}
	where $M_g(t)$ is the time-varying mass and  $M_g(0)$ is the gaseous  mass at birth time. The characteristic timescale, $t_{decay}$,  is given by
	\begin{equation}
		t_{decay}=\frac{r_{hi}}{v_g},
		\end{equation}
	where  $r_{hi}$ is the initial half-mass radius of the cluster and  $v_g=10$ km s$^{-1}$ is the velocity of the heated ionized gas in the  $H_{II}$ region \citep{KroupaArsethHurley2001}.
	
	The stellar IMF is adopted in the form of a three-segment power-law function as follows:
	\begin{eqnarray}
		\xi (m) \propto m^{ -\alpha}: \left\{
		\begin{array}{ll}
			\alpha _1 = 1.35 \hspace{0.25 cm} ,\hspace{0.25 cm} 0.08 <\frac{m}{M_\odot}<0.50\\
			\alpha _2 = 2.35 \hspace{0.25 cm} ,\hspace{0.25 cm} 0.50 <\frac{m}{M_\odot}<1.00\\
			\alpha _3 \hspace{1.25 cm} ,\hspace{0.35 cm} 1.0 <\frac{m}{M_\odot}<100.0
			\end{array} \label{3seg-MF}
		\right.
		\end{eqnarray}
	
	In order to cover the canonical and top-heavy IMF, $\alpha _3$ is varied from 2.3 (canonical Salpeter value) to 1.5.
	%This series of models will allow an extrapolation to larger values of $M_{ecl}$ relevant for GCs to test on the analysis by \cite{MarksKroupaDabringhausen2012}. 
	Table \ref{tab:details-table2} gives an overview of the simulations performed. The retainment of stellar remnants in the cluster is discussed in Section \ref{sec:remnants}. 
	
	It should be noted that,  in the \textsc{nbody6} code and for a given total clusters mass, there are two types of IMF-generating schemes to generate the stellar masses: (i) using the largest value of stellar mass, i.e., 100 $M_{\odot}$, and then continued decreasing of the value according to Equation (\ref{3seg-MF}) until they complete the assumed total cluster mass (optimal sampling, \cite{Kroupa2013}), or (ii) generating stellar masses in a probabilistic fashion until the assumed total mass was reached. We used the latter approach.  Therefore, it is possible to generate at least one star with a maximum mass in the cluster or not.  In Figure \ref{fig:mstar-max-25-95-100}, we compare the number of massive stars in the range from 95 to 100$M_{\odot}$ (top panel) and from 25 to 100$M_{\odot}$ (bottom panel) for models with different initial masses and different values of  $\alpha_3$. The mean value of the generated maximum stellar mass for all models is 99.4$M_{\odot}$, which is very close to the adopted maximum stellar mass in this paper. This implies that the type of IMF-generating schemes of the cluster  has no impact on the results, particularly for the top-heavy IMF. Essentially, for the cluster masses studied here, optimal sampling and random sampling lead to next-to-identical results.

	\section{RESULTS} \label{sec:Results}
	
	At the beginning of the evolution, the mass loss is dominated by gas expulsion and the  evolution of massive stars.  Depending on the abundance of heavy stars in each model, which is determined by $\alpha_3$,  the clusters lose about 30\% to 80\% of their mass within the first 100 Myr through stellar evolution. Here we present the results of the evolution of different  characteristic parameters of the model clusters. Throughout the simulations, we keep only bound stars and remove stars with $r>r_t$ from the computations, where $r_t$ is the instantaneous tidal radius, which is given by

	\begin{equation}
		r_t=R_G\left(\frac{M(t)}{3M_{gal}(\leq R_G)}\right)^{1/3},
		\end{equation}
	where $R_G=8.5$ kpc, $M_{gal}$ is the mass of the hosting galaxy within $R_G$, and $M(t)$ is the mass of the cluster at time $t$, which is calculated from summing over all stars bound to the cluster.

	\subsection{Evolution of the mass and the half-mass radius}
	
	Using the models described in the previous section, we now discuss the evolution of the mass and the size of clusters,  starting with different degrees of top-heaviness of the IMF and undergoing early gas expulsion.  In order to understand how the top-heaviness of the IMF (i.e., the value of $\alpha_3$) influences  the half-mass radius of bound stars  and the amount of mass lost by a star cluster during the evolution, we first consider initially identical star clusters but with a different degree of the top-heaviness ( summarized in  Table \ref{tab:details-table2}).
	
	The evolution of the total masses and half-mass radii of stars bound to the clusters is shown in Figure \ref{fig:m-rh-time-fig1}. As a result of the dilution of the gravitational potential well that is driven by the early gas expulsion and rapid mass loss due to the evolution of the massive stars,  all models expand violently and lose a fraction of their stars depending on their stellar MF slope (Figure \ref{fig:m-rh-time-fig1}). This early ejection of gaseous mass leads to a positive mean radial velocity  of the  stars, and hence a large number of stars move outward. Stars with sufficiently high energies become unbound from the cluster and set the amount of mass lost. Less energetic stars migrate outward while remaining bound, thus driving the expansion of the cluster.
	
	Three mass-loss phases can be recognized in the evolution in Figure \ref{fig:m-rh-time-fig1}. In the first phase, which is shown as a rapid drop in the mass of the cluster, the mass loss is dominated by gas expulsion and stellar evolution. Mass loss due to the stellar evolution strongly depends on the  $\alpha_3$ parameter. In the second phase, mass loss is dominated by internal and external dynamical effects. The third phase is after core collapse where  the mass-loss rate is higher than before core collapse.

	The mass-loss rate is higher for the cases with a more top-heavy IMF.  
	Clusters that initialize with a more top-heavy IMF will show a larger maximum value for the half-mass radius. The cluster loses about 80\% of its mass by early rapid gas expulsion and stellar evolution in the case of the most top-heavy IMF (with $\alpha_3=1.5$), but this value decreases to 40\% for clusters with a canonical IMF (with $\alpha_3=2.3$). Consequently, the clusters with a top-heavy IMF dissolve faster than the clusters with a canonical  IMF. As can be seen, for $M_{ecl} = 10^4 ~ M_\odot$ (top left panel) all of the clusters will be dissolved before 6 Gyr. But massive clusters with $\alpha_3 > 1.9$ will survive longer than 12 Gyr.  
	
	All models with the same mass and half-mass radius evolve on the same circular orbit with a galactocentric distance of $R_G=8.5$ kpc, such that the mass loss due to tidal stripping  depends on  $\alpha_3$. This is because the expansion rate of each model is a function of $\alpha_3$ (see Table \ref{tab:fitting-param-table1} for details).   Therefore,  even for tidally underfilling clusters with an initially top-heavy IMF,  the early dramatic expansion  makes the clusters become tidally filling and therefore more affected  by tidal stripping. 
	
	In Figure \ref{fig:rh-horizontal-5Myr-fig2}, we zoom in on the time evolution of the half-mass radii in the first 20 Myr for two representative computed $N$-body models with different initial mass. The half-mass radius of our models for the first 20 Myr of cluster evolution has a similar evolutionary trend to that found by \citet{Brinkmann2017} and \citet{KroupaArsethHurley2001} for the evolution with a canonical IMF. At the beginning, the half-mass radii remain fixed throughout the cluster evolution until $t_D=0.6$ Myr (the ultracompact $H_{II}$ phase). At time $t >t_D$ the gas mass evolves according to Equation (\ref{equ:gas-mass}). At a cluster age around\footnote{According to the SSE package that is used here (see Section \ref{sec:Initial Conditions} for more details), $t=3.5$ Myr is the typical lifetime of a 100  $M_{\odot}$ star.} $t=3.5 Myr$, stellar evolution and Type II supernova explosions cause a noticeable mass loss, and hence models with different values of $\alpha_3$  separate at this time of their evolution. As  expected, the more top-heavy IMF models can expand faster because of the larger population of massive stars.

	\subsection{Dissolution time}
	
	For clusters surviving the early evolutionary stages, mass loss due to the  tidal field from the  host galaxy and two-body relaxation are the main processes determining their dissolution time.  Moreover, these processes affect the cluster's contribution to the host galaxy's field stellar population.  
	
\begin{figure}[]
		\centering
		\includegraphics[width=87mm]{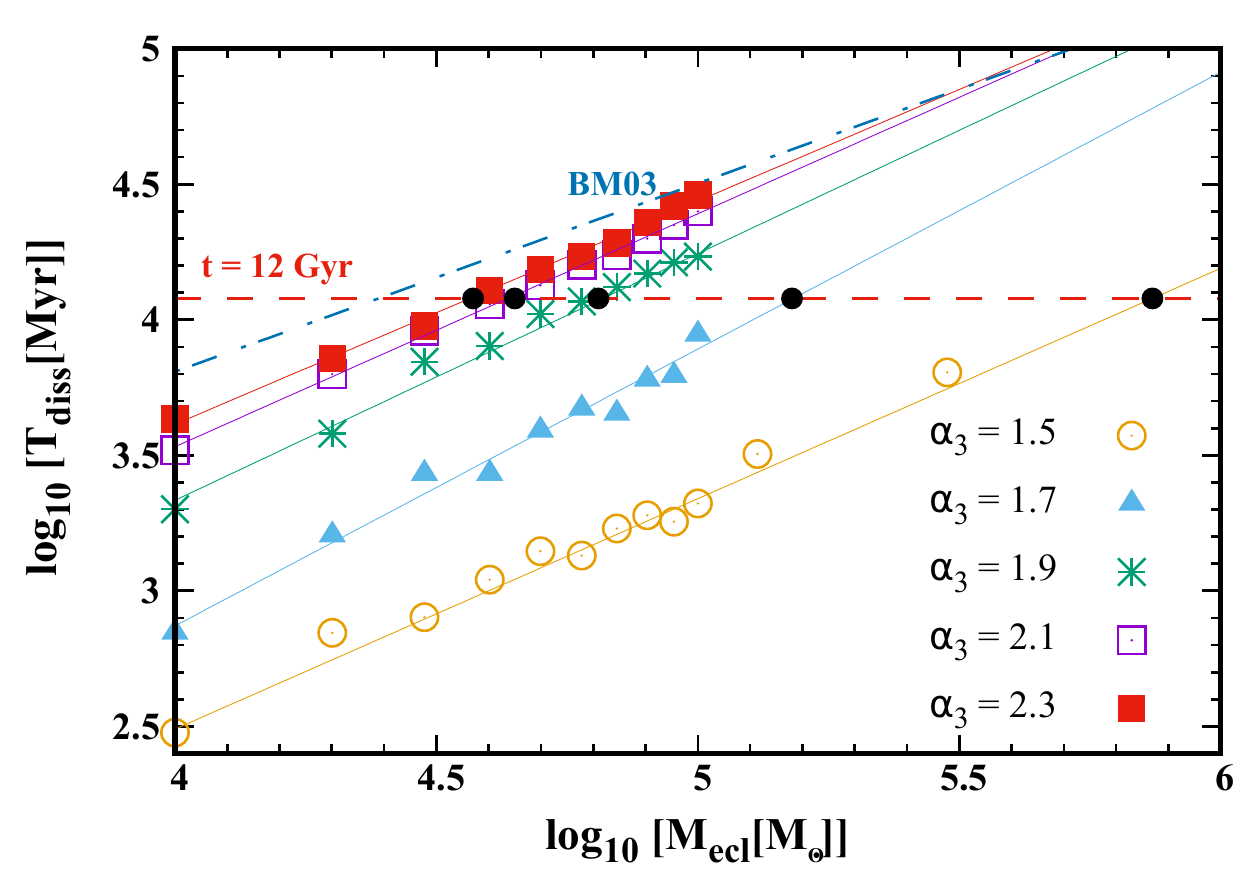}
		\caption{Dissolution time vs. initial  mass of embedded clusters for different values of $\alpha_3$. The points show the results of $N$-body simulations for a given initial mass and $\alpha_3$  until dissolution time. The horizontal dashed line indicates $t=12$ Gyr. The clusters that are evaporated before $t=12$ Gyr are shown below this line, while the surviving clusters are above the horizontal line. The filled black circles show the minimum initial  mass of the clusters for them to survive for each $\alpha_3$ (Table \ref{tab:fitting-param-table1}). The rising solid lines show linear fits to the results of the simulations  for each $\alpha_3$ (Equation (\ref{equ:tdiss})). The fitting parameters are given in Table \ref{tab:fitting-param-table1}. The dotted-dashed blue line shows the fit to the results of \citet{BaumgardtMakino2003} with a canonical IMF and without residual gas expulsion.}
		\label{fig:tdiss-mass-all-fig3}
		\end{figure}

	Without considering the effect of residual gas expulsion, \citet{BaumgardtMakino2003} showed that the lifetime of star clusters moving on similar orbits scales with the relaxation time as $ T_{diss} \propto T_{rh}^x$, where the exponent $x$ depends on the initial concentration of the cluster and is around $x \approx 0.8$. Figure \ref{fig:tdiss-mass-all-fig3}  depicts the cluster lifetime, $T_{diss}$, as a function of the initial mass in stars of the embedded cluster, $M_{ecl}$, for various $\alpha_3$.   As can be seen, the dissolution time of clusters increases linearly with $M_{ecl}$.
	
	However, even though the strength of the tidal gravitational interaction due to the host galaxy is the same for both canonical and top-heavy IMFs, dissolution is faster for clusters with the more top-heavy IMF (i.e. lower $\alpha_3$). This is due to the larger amount of  mass loss from stellar evolution from clusters with the more top-heavy IMF and rapid gas expulsion, which leads to a strong expansion.  When the clusters expand, they change from an initially tidally  underfilling phase into the tidally filling phase, where by truncating the cluster sizes the  tidal field significantly enhances the mass-loss rate.  
	
\begin{figure}[]
		\centering
		\includegraphics[width=70mm]{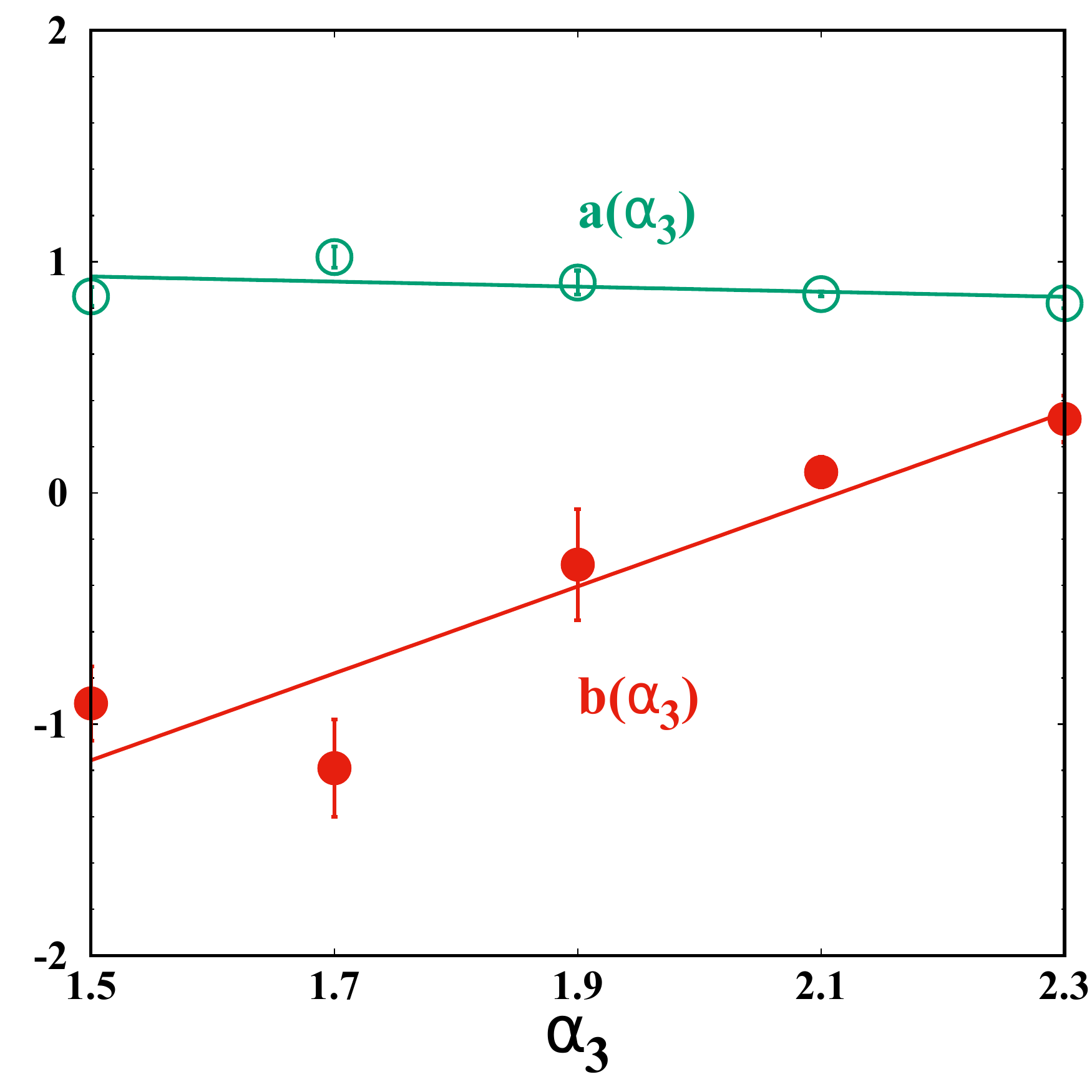}
		\caption{Linear-fit coefficients of Equation (\ref{equ:tdiss}) as a function of $\alpha_3$. The open circles show the slope of the linear fit to $T_{diss}$ that are calculated as a result of the linear fit in Figure \ref{fig:tdiss-mass-all-fig3} as a function of $\alpha_{3}$. The filled circles depict the vertical intercept of Equation (\ref{equ:tdiss}).  The solid lines are the linear fit to these shown data ( Equations (\ref{equ:a-tdiss})  and (\ref{equ:b-tdiss})).}
		\label{fig:a-b-mmin-fit-fig4}
		\end{figure}

	The initial degree of top-heaviness therefore determines the dependence of $T_{diss}$ on $M_{ecl}$.  We found that the dissolution time can be written as
	\begin{equation}
		\rm \log_{10} \left[T_{diss}(\alpha_3) \right]=a(\alpha_3)~\log_{10} \left[ M_{ecl} \right]+b(\alpha_3), \label{equ:tdiss}
		\end{equation}
	where the coefficients  $a(\alpha_3)$ and $b(\alpha_3)$ depend on the degree of top-heaviness.  For clusters with $\alpha_3=1.5$, the coefficient is $a=0.85\pm 0.04$, and the canonical models with $\alpha_3=2.3$ have almost the same dependence on $R_G$, with  $a=0.82\pm 0.02$. For all models, the coefficient $a(\alpha_3)$ is the same  within the 2$\sigma$ error bar and is nearly independent of the $\alpha_3$ index (Table \ref{tab:fitting-param-table1}). However, the offset in $T_{diss}$ is large. It is therefore not possible to derive a unique power law to fit the scaling of $T_{diss}$ with $R_G$ for all values of the $\alpha_3$ index. In other words, for top-heavy IMF models, the dissolution time is significantly shorter than for systems with a canonical IMF. The linear best-fit parameters for the $T_{diss}-M_{ecl}$ relation for each value of $\alpha_3$ are listed in Table \ref{tab:fitting-param-table1}.  
	In Figure \ref{fig:a-b-mmin-fit-fig4} we plot  the best-fit values of $a$ and $b$ as a function of $ \alpha_3$. The following functions $a(\alpha_3)$ (green line) and $b(\alpha_3)$ (red line) are linear fits to the data in Table \ref{tab:fitting-param-table1}:  
	\begin{equation}
		a(\alpha_{3})=(-0.11\pm  0.13)~\alpha_{3}+(1.10\pm  0.25),
		\label{equ:a-tdiss}
		\end{equation}
	%%1.101 - 0.11 x
	\begin{equation}
		b(\alpha_{3})=(1.88\pm  0.47)~\alpha_{3}-(3.98 \pm 0.91).
		\label{equ:b-tdiss}
		\end{equation}
	Therefore, for any system with a top-heavy IMF  it is possible to find out the  dissolution time by Equations (\ref{equ:tdiss}) - (\ref{equ:b-tdiss}).

	The slope of our $T_{diss}-M_{ecl}$ relation for canonical IMF models is reasonably close to that of \cite{BaumgardtMakino2003}, who found that $T_{diss}\propto M_{ecl}^x$. The difference may well be due to the different initial conditions that are assumed in both studies. We assume the Plummer model for the mass distribution, while \cite{BaumgardtMakino2003} set up a King model with $W_0=7.0$. Moreover, the canonical IMF with an upper and lower mass limit of 15 and 0.1 $ M_{\odot}$, respectively, has been assumed by \cite{BaumgardtMakino2003}, while in our set the maximum and minimum stellar masses are 100 and 0.08 $ M_{\odot}$, respectively. Maybe more importantly, however, \cite{BaumgardtMakino2003} investigated clusters without early gas expulsion, whereas our clusters start with a rapid gas expulsion.

	%%%%%%%%%%

	%%%%%%%%%%%%%%%%%
	
	%%%%%%%%%%%%%%%%%
	\begin{figure}[]
		\centering
		\includegraphics[width=8cm]{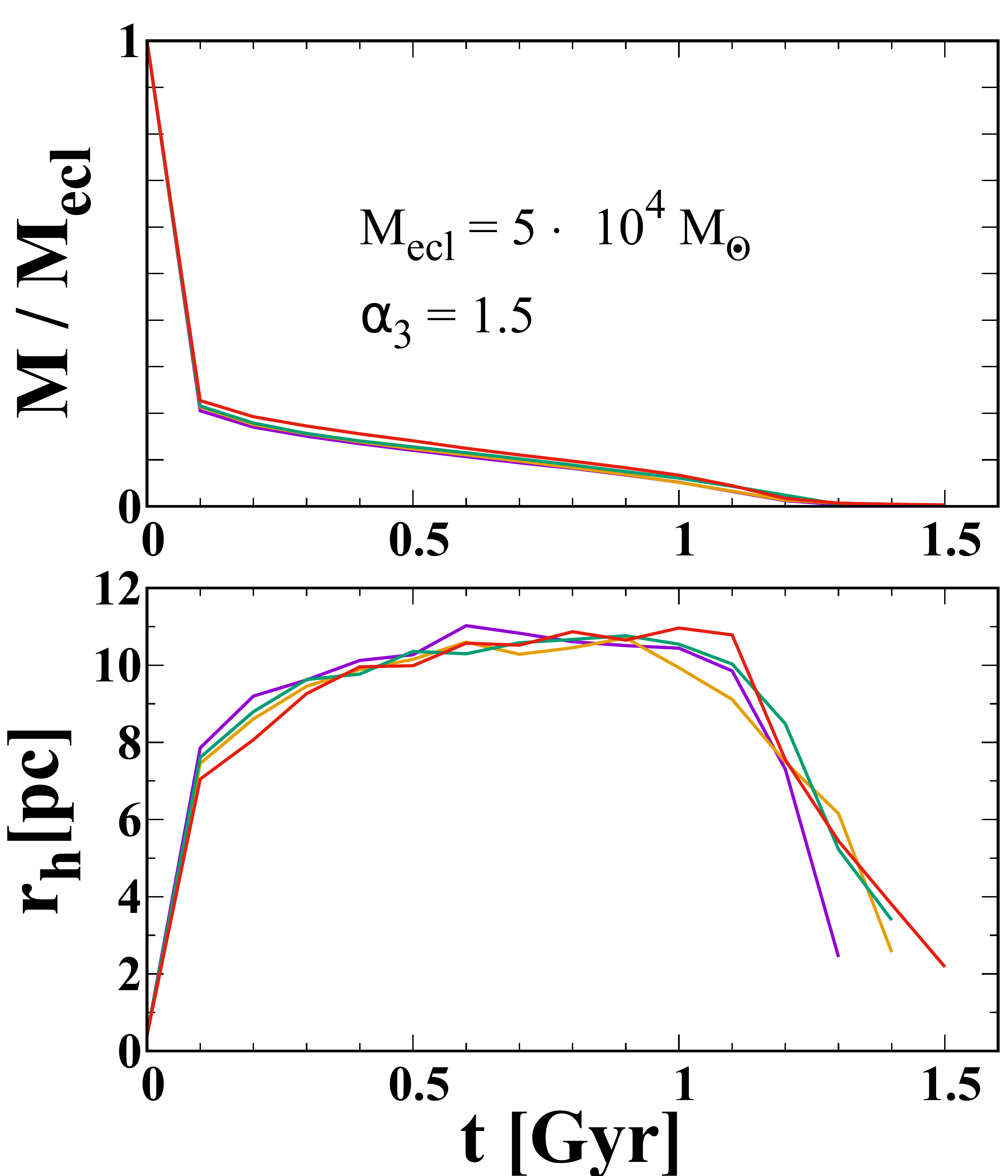}
		\caption{ Comparison of the mass-loss rate and the evolution of the half-mass radius for four simulations with the same initial half-mass radius of 0.4 pc, the initial mass of $5 \times 10^4 M_{\odot}$ and  with $\alpha_3=1.5$, but with different initial random number seeds for generating the initial velocity and position of stars. The typical error of the numerical values of the resulting parameters is about 5\%.  }
		\label{fig:different-random-seeds}
		\end{figure}
	
	%%%%%%%%%%

	\begin{figure}[h]
		\centering
		\includegraphics[width=80mm]{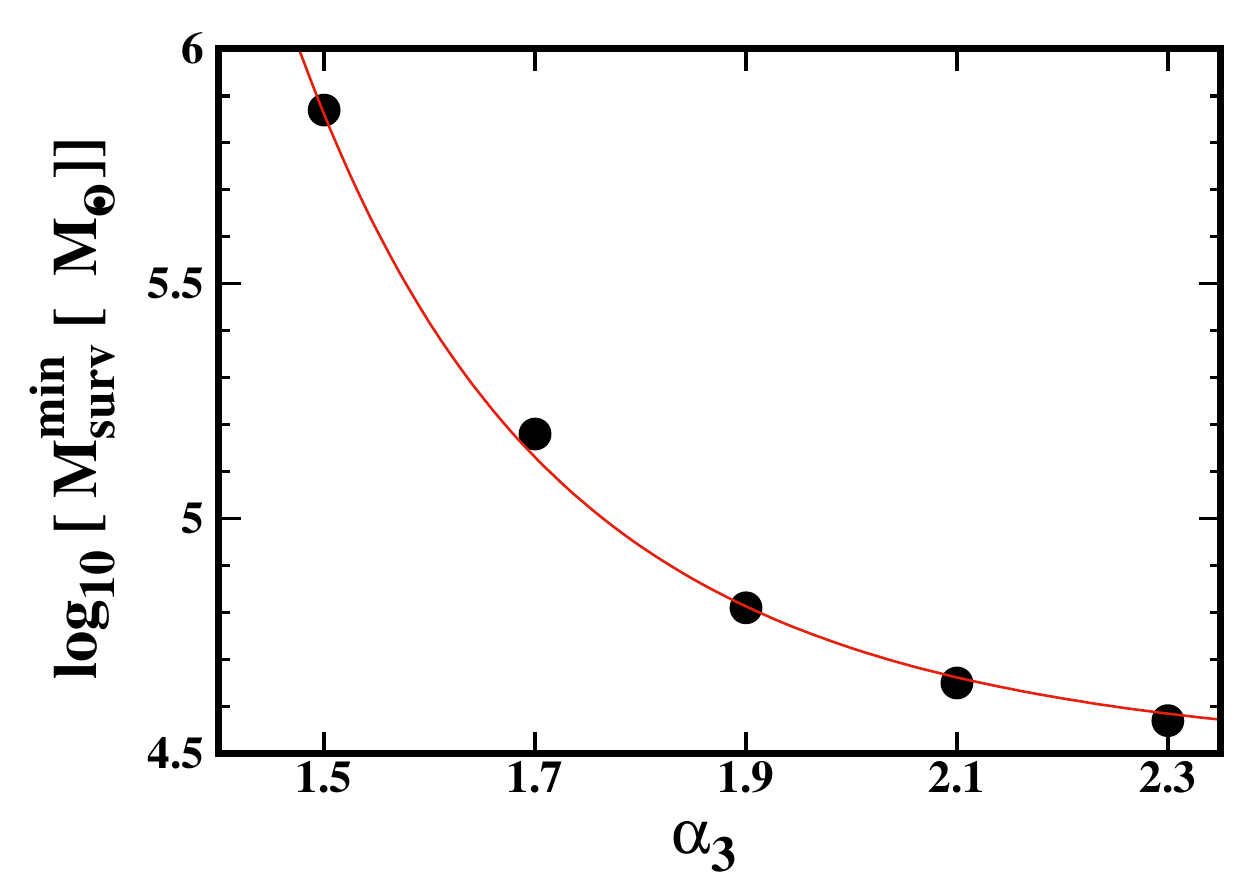}
		\caption{Minimum surviving mass of embedded  clusters, $M^{min}_{surv}$ for various $\alpha_3$. The filled circles show the minimum mass of embedded clusters for every $\alpha_3$ such that clusters heavier than these points will survive longer than 12 Gyr. The solid line is the best-fitted curve in the form of Equation (\ref{equ:minimum-survived-fit}). }
		\label{fig:Minimum survived mass-fig6}
		\end{figure}

	%%%%%%%%%%%%%%%%%%%%%%%%%%%%%%%%%%%%%%%%%%%%%%%%%%%%%%%%%%%% %%%%%%%%%%%%%%%%%%%  NEW
	
	It should be noted that there is statistical dispersion in the results if we repeat the calculations for one arbitrary model with the same initial parameters but with a different random seed. Because of the computational cost of direct $N$-body simulations of the evolution of dense star clusters with small initial half-mass radii (it takes a few weeks for one simulation to complete even on GPU-based computers), we are limited to a single simulation per model. Therefore, statistics for all parameters by repeating several runs for each simulation are CPU compromised at this time.

	Here, we examine how changing the initial random seed for generating the positions and velocities of each star influences the evolution of  mass, half-mass radius, and dissolution time. This will allow us to estimate the dispersion and hence the vertical error bars in Figure \ref{fig:tdiss-mass-all-fig3}. We carried out four further simulations  for a particular model with the same initial conditions but with different random seeds for generating the initial velocities and positions of the stars. Figure \ref{fig:different-random-seeds} shows the mass loss and the evolution of the half-mass radius for four simulations with an initial half-mass radius of 0.4 pc and an initial mass of 50,000 $M_{\odot}$  (corresponding to the initial particle number of $N\approx 24600$), both of them with identical initial conditions but with different random seeds. The dissolution times of these models are 1.40, 1.30, 1.40, and 1.50 Gyr.  The mean central value and dispersion of the dissolution time of these four clusters are $1.40 \pm 0.07$ Gyr, i.e., the uncertainty of the resulting parameters is about 5\%.

	\setcounter{table}{0}
	\begin{table}[]
		\renewcommand{\thetable}{\arabic{table}}
		\centering
		\caption{Best-fitting Coefficients for the  $T_{diss}-M_{ecl}$ Relation  for Different values of $\alpha_3$. The first column gives the Values of $\alpha_3$. $a$ and $b$ are the slope and the vertical intercept of the fitted line  (Equation (\ref{equ:tdiss})) for various $\alpha_3$.  The last column gives the minimum surviving embedded mass of clusters, $M^{min}_{surv}$, after 12 Gyr for different values of $\alpha_3$. The corresponding value for the slope derived by \citet{BaumgardtMakino2003} is given in the bottom line. } 
		\label{tab:fitting-param-table1}
		\begin{tabular}{cccc}
			\tablewidth{0pt}
			\hline
			\hline
			$\alpha _ 3$ & $a(\alpha_3)$ & $b(\alpha_3)$ & $M^{min}_{surv}$ \\
			&  &  & $[10^3 M_\odot]$\\
			\hline
			\decimals
			1.5 & 0.85 $\pm$ 0.04 & -0.91 $\pm$ 0.16  & 700  \\%1.05*x-1.8
			1.7 & 1.01 $\pm$ 0.05 & -1.19 $\pm$ 0.21  & 150  \\%1.0213*x-1.2144
			1.9 & 0.91 $\pm$ 0.05 & -0.31 $\pm$ 0.24  & 65 \\ %0.91*x-0.306
			2.1 & 0.86 $\pm$ 0.01 & 0.09  $\pm$ 0.07  & 45  \\%0.85935233*x+0.094341016
			2.3 & 0.82 $\pm$ 0.02 & 0.32  $\pm$ 0.10  & 38  \\ %0.8236923*x+0.319572730
			\hline
			2.3*& 0.82* & ---  & --- \\
			\hline
			\multicolumn{4}{c}{  *Denotes the value by BM03}
			\end{tabular}
		\end{table}

	%%%%%%%%%%%%%%
	
	\subsection{Minimum surviving mass}
	
	The influence of the top-heaviness of the IMF on the evolution and survival/destruction of star clusters is examined here by  obtaining a relation for the minimum surviving mass of the cluster ($M^{min}_{surv}$, i.e., the minimum $M_{ecl}$ that survives within a Hubble time) as a function of $\alpha_3$. The horizontal dashed line in Figure \ref{fig:tdiss-mass-all-fig3} shows the time $=$ 12 Gyr.  The intersection of the $T_{diss}-M_{ecl}$ relation for each value of $\alpha_3$  with this horizontal line (black circles) specifies the corresponding $M^{min}_{surv}$ (see Table \ref{tab:fitting-param-table1} for the values).  Determining  $M^{min}_{surv}$ might help to understand the contribution of dissolved star clusters to different Galactic components.

	%Figure \ref{Minimum survived mass-fig7 } shows that the impact of top-heaviness becomes less pronounced with increasing the .... due to .....
	
	Figure \ref{fig:Minimum survived mass-fig6} shows the minimum surviving mass (listed in Table \ref{tab:fitting-param-table1}) for different values of $\alpha_3$.  With the importance of $\alpha_3$ implied by Figure \ref{fig:tdiss-mass-all-fig3}, the required minimum initial cluster mass to survive for a  Hubble time increases for clusters with a more top-heavy IMF.   Moreover, by a polynomial least-squares fit to the data we found that the minimum surviving mass scales  with $\alpha_3$ as

	\begin{equation}
		\log_{10}(M^{min}_{surv}/M_\odot)=A~(\alpha_3)^{-\eta}+B,
		\label{equ:minimum-survived-fit}
		\end{equation}
	%4.33189 + 6.54797/x^4
	where the coefficients $A=15.7  $ , $\eta=6$ and $B=4.5 $.   Therefore, for a given value of $\alpha_3$  one can calculate the mass of the embedded cluster that survives for a Hubble time. For example, when $\alpha_3=1.5$ clusters at $R_G=8.5$ kpc heavier than $7\times 10^5 M_{\odot}$ will survive longer than 12 Gyr,  while for the canonical IMF, all clusters with a mass lower than $4\times 10^4 M_{\odot}$ will be dissolved.

	\begin{figure}[]
		\centering
		\includegraphics[width=9cm]{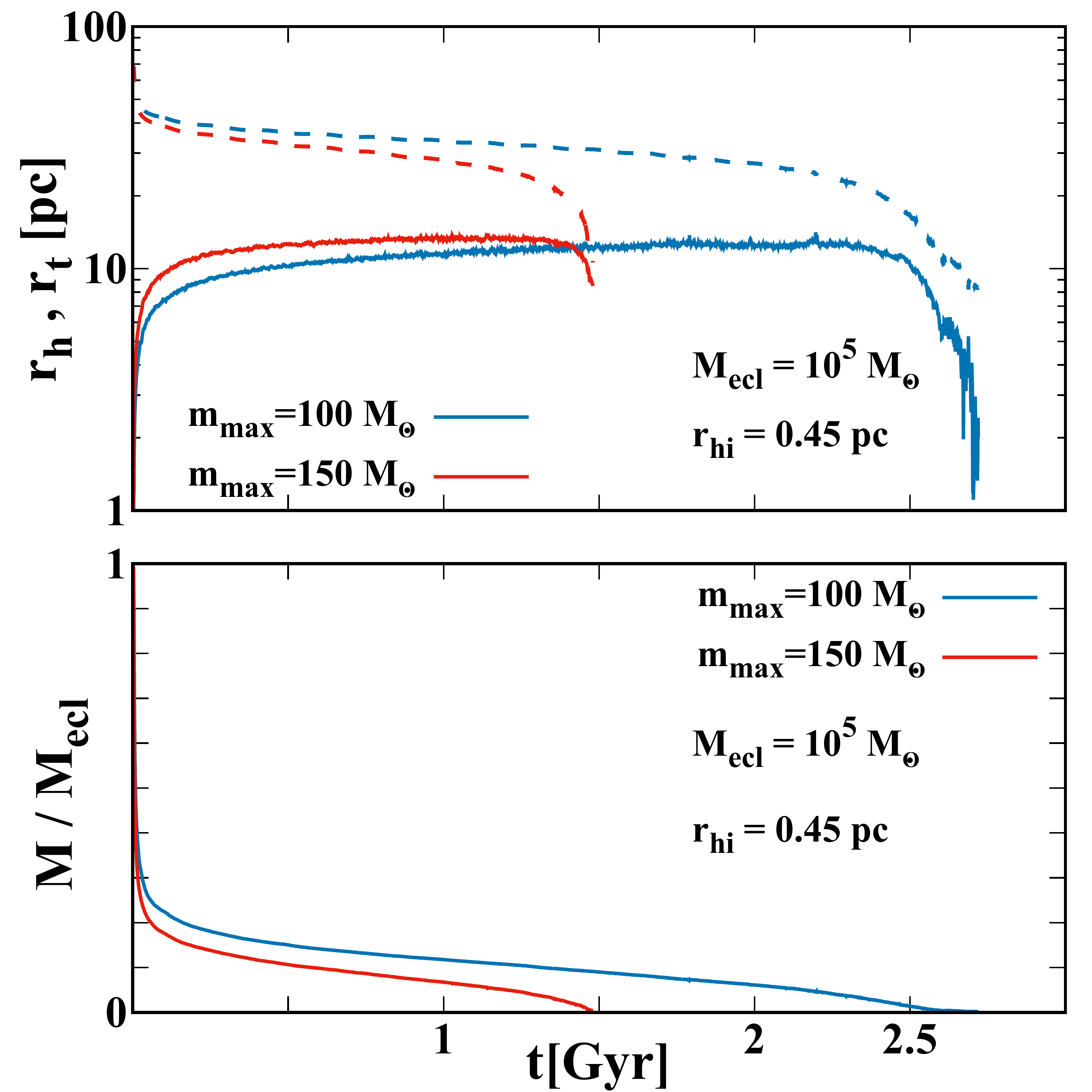}
		\caption{Comparison of the half-mass radius, tidal radius and the total mass for two clusters with the same initial conditions but different upper stellar limits of stellar mass. The half-mass radius (solid line) of the cluster that has an upper stellar mass of  $m_{max}=150 M_{\odot}$ expands more and dissolves faster than the other one in the top panel. The tidal radius (dashed line) is decreasing for both models. }
		\label{fig:mmax-150-100}
		\end{figure}
	
	\begin{figure}[!htb]
		\centering
		\includegraphics[width=9cm]{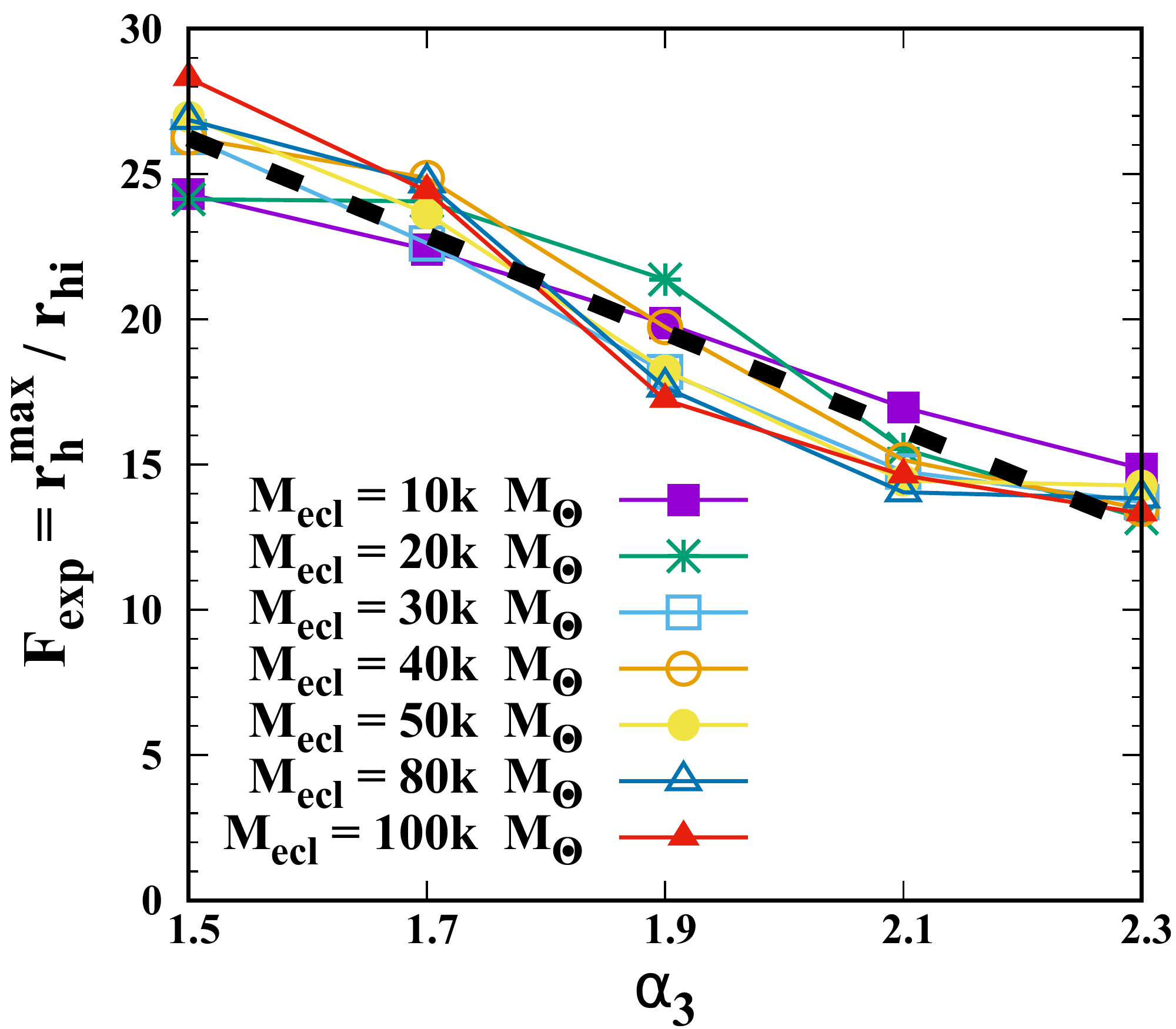}
		\caption{Expansion factor (Equation (\ref{equ:expansion})) is plotted for different $\alpha_3$. The clusters with an initially top-heavy IMF generally expand stronger than those with a canonical IMF. The largest expansion factors are around 25 - 30 for $\alpha_3=1.5$. The linear fit to the data is plotted as a dashed line (see Equation (\ref{equ:expansion})). }
		\label{fig:expansion-fig7}
		\end{figure}

	\begin{figure*}[t]
		\centering
		\includegraphics[width=8.9cm]{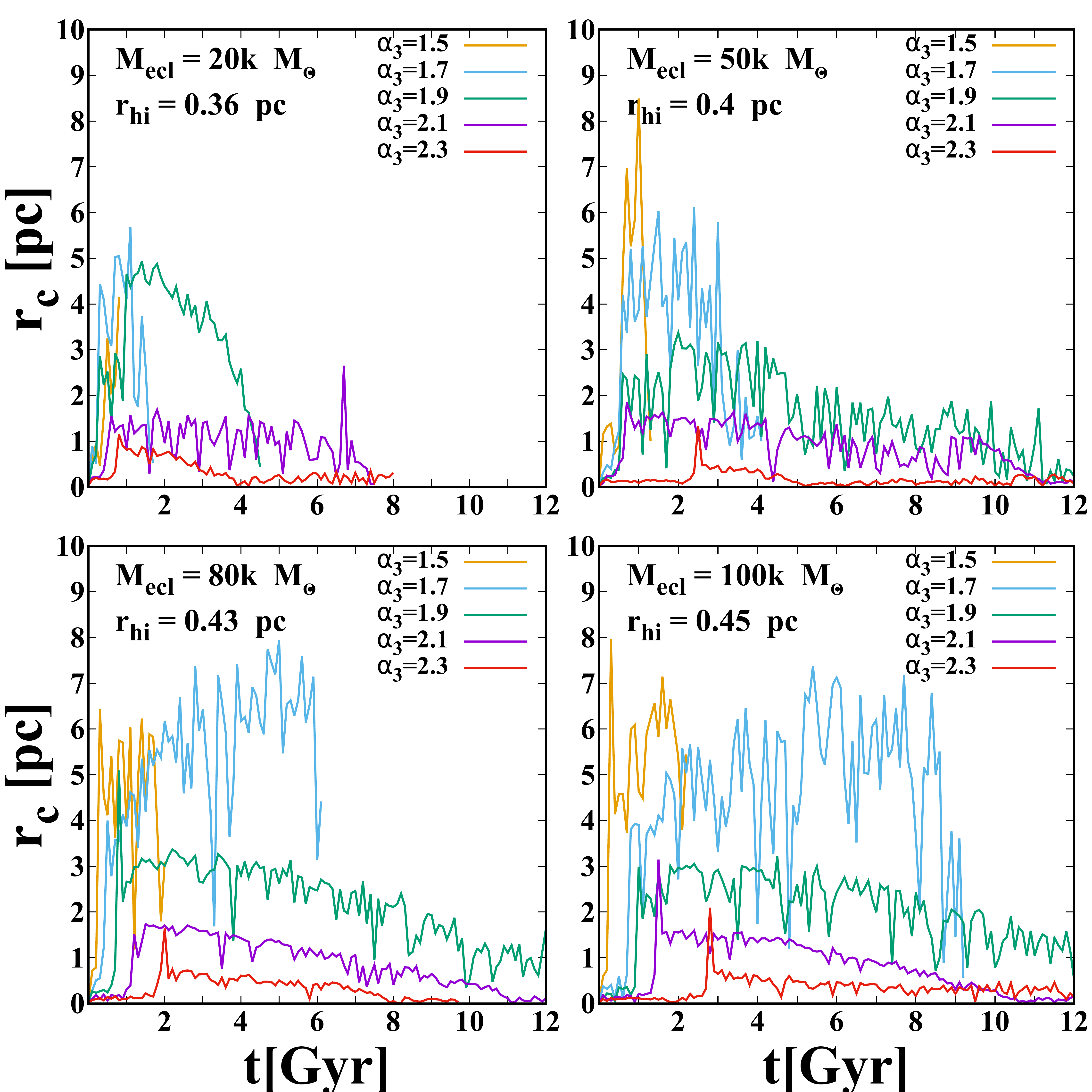}
		\includegraphics[width=8.9cm]{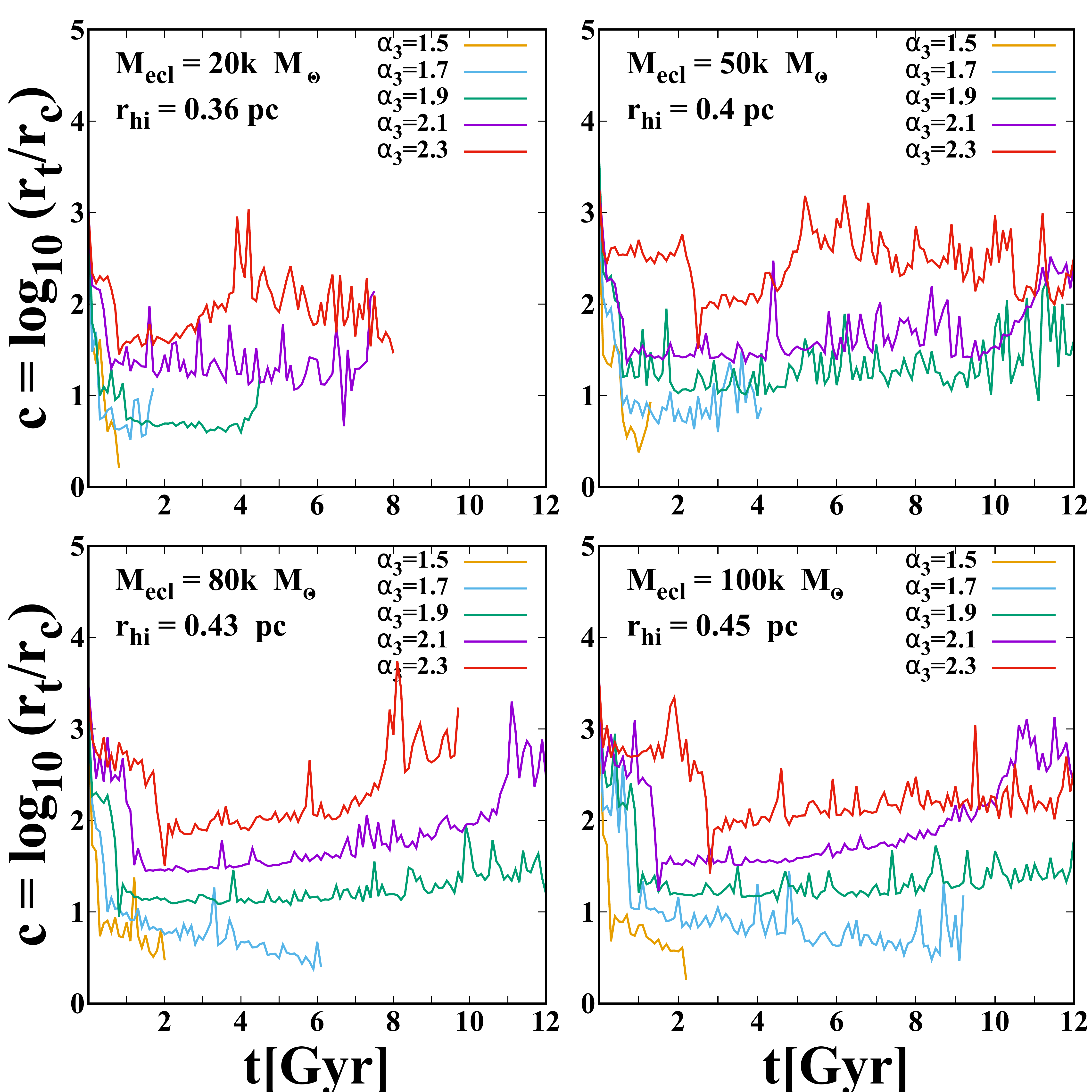}
		\caption{Evolution of the core radius and the central concentration parameter, $c=\rm log_{10}[r_t/r_c]$ are plotted as a function of time for models with $M_{ecl}=$ 20k, 50k, 80k, and 100k $M_\odot$. In all panels the value of core radius is smaller for the models with the canonical IMF, and consequently their $c$-value is higher than for the case of the top-heavy IMF. }
		\label{fig:rc}
		\end{figure*}

	\subsection{The impact of the maximum stellar mass}
	
	Here we examine how changing the  maximum stellar mass of a star cluster (that undergoes gas expulsion), which determines the amount of stellar mass that is lost within the first few Myr of a cluster's lifetime, influences its evolution and dissolution time. 
	
	Figure \ref{fig:mmax-150-100} shows the  evolution of the mass and half-mass radius for two simulations  with identical initial conditions but with different values of the upper mass limit of the stellar IMF.  We consider two different models with $m_{max}=100$ and $150 M_{\odot}$ and an initially top-heavy IMF ($\alpha_3=1.5$).  Both models are evolving at a galactocentric distance of $R_G=8.5$ kpc, with the same initial mass of $M_{ecl}=10^5 M_{\odot}$ and an initial half-mass radius of $r_{hi}=0.45$ pc. 
	
	As can be seen in Figure \ref{fig:mmax-150-100}, the early expansion is stronger in models with $m_{max}=150 M_{\odot}$ owing to the larger number of heavy stars, but the cluster survives for up to 1.5 Gyr. Since both clusters are  tidally underfilling, they expand owing to gas expulsion and reach  significantly larger half-mass radii to become tidally filling within about $100$ Myr. However, the half-mass radius of the model with $m_{max}=150 M_{\odot}$ reaches a larger value than the model with $100 M_{\odot}$. The difference between the two half-mass radii is about 20\%. Since the long-term cluster mass loss for these clusters can be regarded as a runaway overflow over the tidal boundary, the model with $150 M_{\odot}$ disrupts faster than the model with $m_{max}=100 M_{\odot}$.

	\subsection{Expansion}

	Here we investigate the dependence of the expansion factor, which is defined as 
	\begin{equation}
		F_{exp} = r^{max}_{h} / r_{hi},   
		\end{equation} \label{exp-factor}
	on the top-heaviness of the IMF. Figure \ref{fig:expansion-fig7} shows the expansion factor as a function of $\alpha_3$. The data points are taken from Table \ref{tab:details-table2}.  Figure \ref{fig:expansion-fig7}  shows that the expansion of a cluster does not depend on its initial embedded cluster mass, $M_{ecl}$.  
	Depending on the top-heaviness of the IMF, the maximum half-mass radii of the clusters expand  by a factor of $F_{exp}=15-30$. While clusters with a canonical IMF ($\alpha_3=2.3$) undergo an expansion by a factor of $F_{exp}=15$, the most top-heavy models expand by up to a factor of $F_{exp}=30$.

	% CONCLUSION or ABSTRACT: ... 
	In order to find the dependence of the expansion factor on the top-heaviness of the IMF, we fit the  expansion factor by a linear formula as follows:   
	\begin{equation}
		F_{exp}(\alpha_3)= e~ (\alpha_3 ) + f,
		\label{equ:expansion}
		\end{equation}
	where the coefficients are determined from a least-squares fit to the $N$-body results. The best-fitted line with  $e=-16.7 \pm 1.7 $ and $f = 51.3 \pm 3.3$ is shown as the dashed black  line in Figure \ref{fig:expansion-fig7}.  
	
	%%%%%%%%%%%%%%%%%%%%%%%%%%%%%%%%%%%%%%%%%%%%%%%   Concentration     %%%%%%%%%%%%%%%%%%%%%%%%%%%%%%%%%%%%%%

	\begin{figure}[t]
		\centering
		\includegraphics[width=8.9cm]{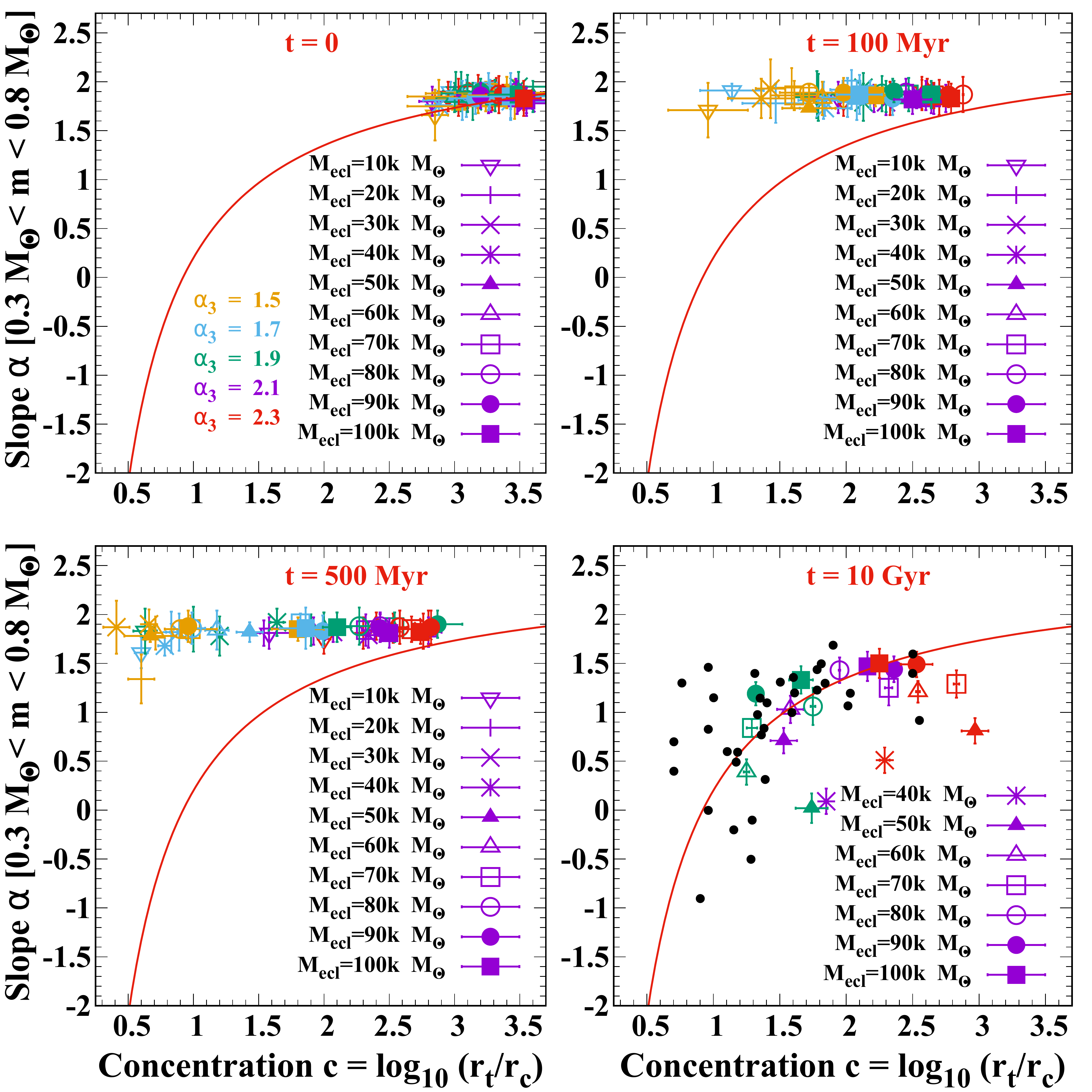}
		\caption{MF slope, $\alpha$, for stars in the mass range  0.3 - 0.8 $M_{\odot}$  as a function of the central concentration parameter. Each point represents one simulated cluster model, with the color indicating the cluster's $\alpha_3$ value. The yellow, blue, green, purple and red symbols correspond to models with  $\alpha_3=$ 1.5, 1.7, 1.9, 2.1, and 2.3, respectively. The red solid line shows the $\alpha-c$ relation from \cite{Demarchi2007}. The calculations are for stars that are located inside the tidal radius. The black filled circles show the data taken from \cite{Paust10} and \cite{Demarchi2007}. }
		\label{fig:alpha-concentration}
		\end{figure}
	
	%The black symbols show the results of simulations at the birth time in the left panel.
	%%%%%%
	%%%%%%
	\begin{figure}[t]
		\centering
		\includegraphics[width=8.8cm]{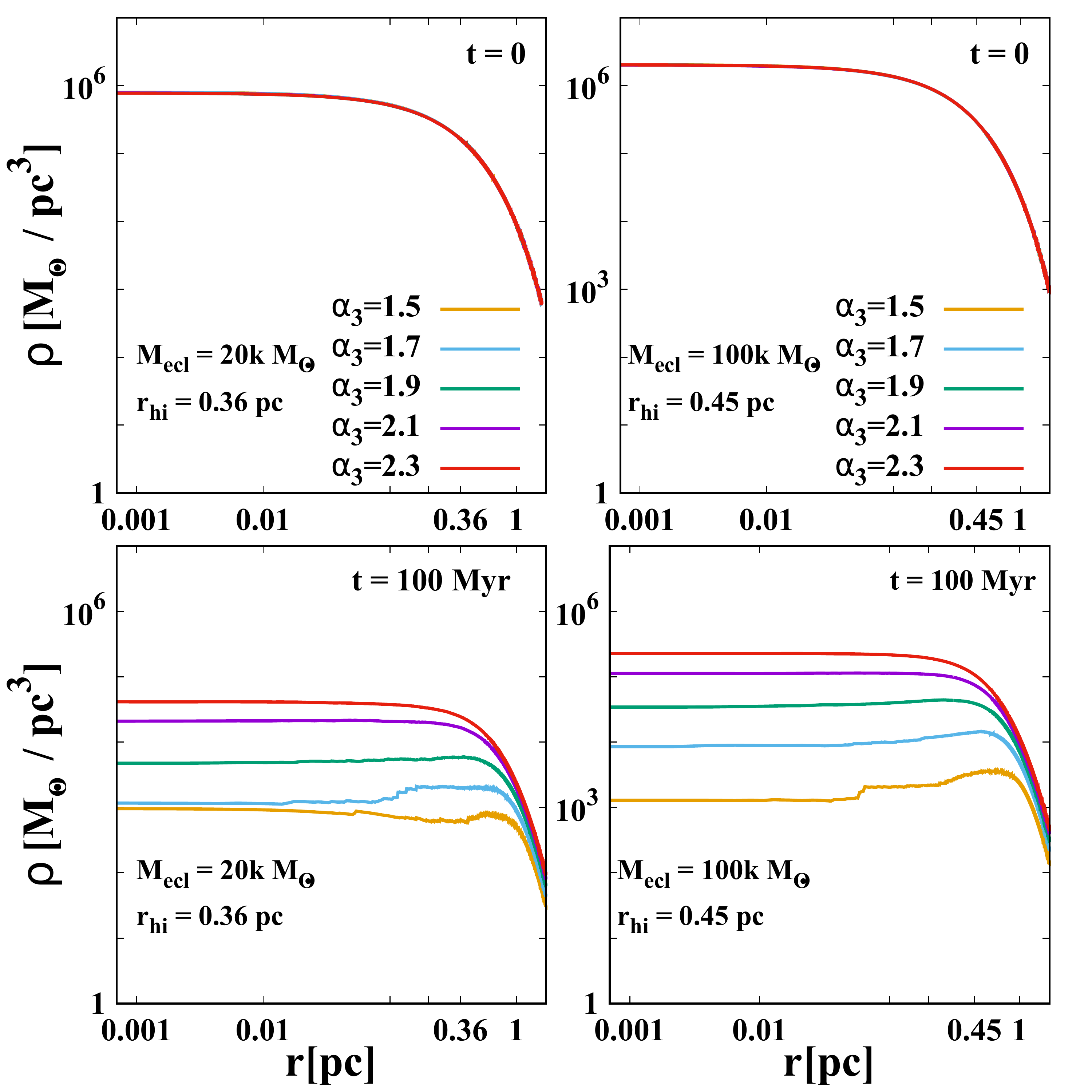}
		\includegraphics[width=8.8cm]{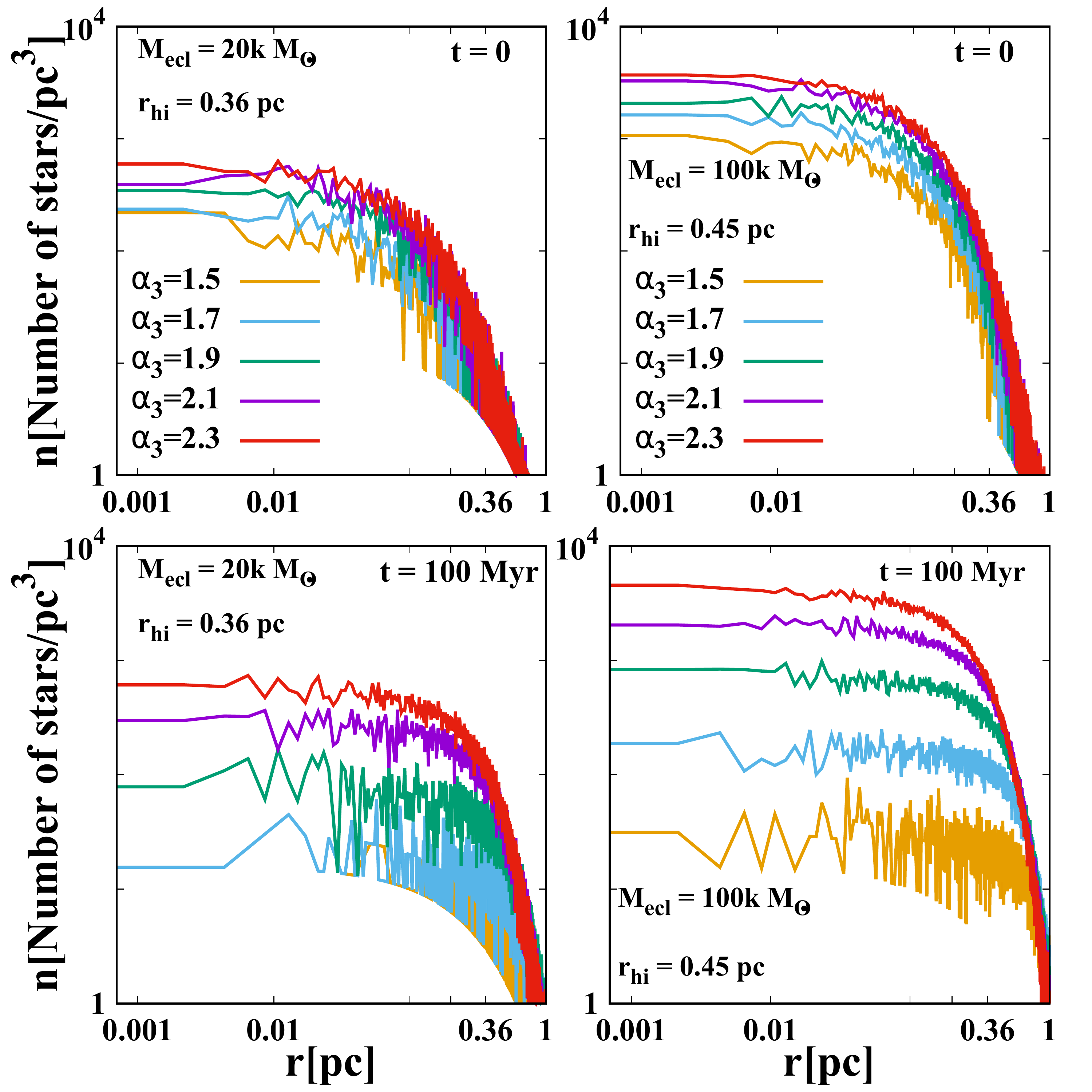}
		\caption{Two snapshots ($t=$0 and 100 Myr) of the mass (upper four panels) and the number (lower four panels) density profile  as a function of distance from the center of the cluster for two selected models with initial masses of 20k and 100k $M_\odot$. Different colors correspond to different values of  $\alpha_3$. In all panels, the mass and number densities in clusters with a canonical IMF are higher than in those with a top-heavy IMF. Note that clusters born with a top-heavy IMF develop a shell structure, with the inner region being less dense than the shell.}
		\label{fig:density-fig9}
		\end{figure}
	%%%%%
	%%%%%
	
	\begin{figure}[t]
		\centering
		\includegraphics[width=8.9cm]{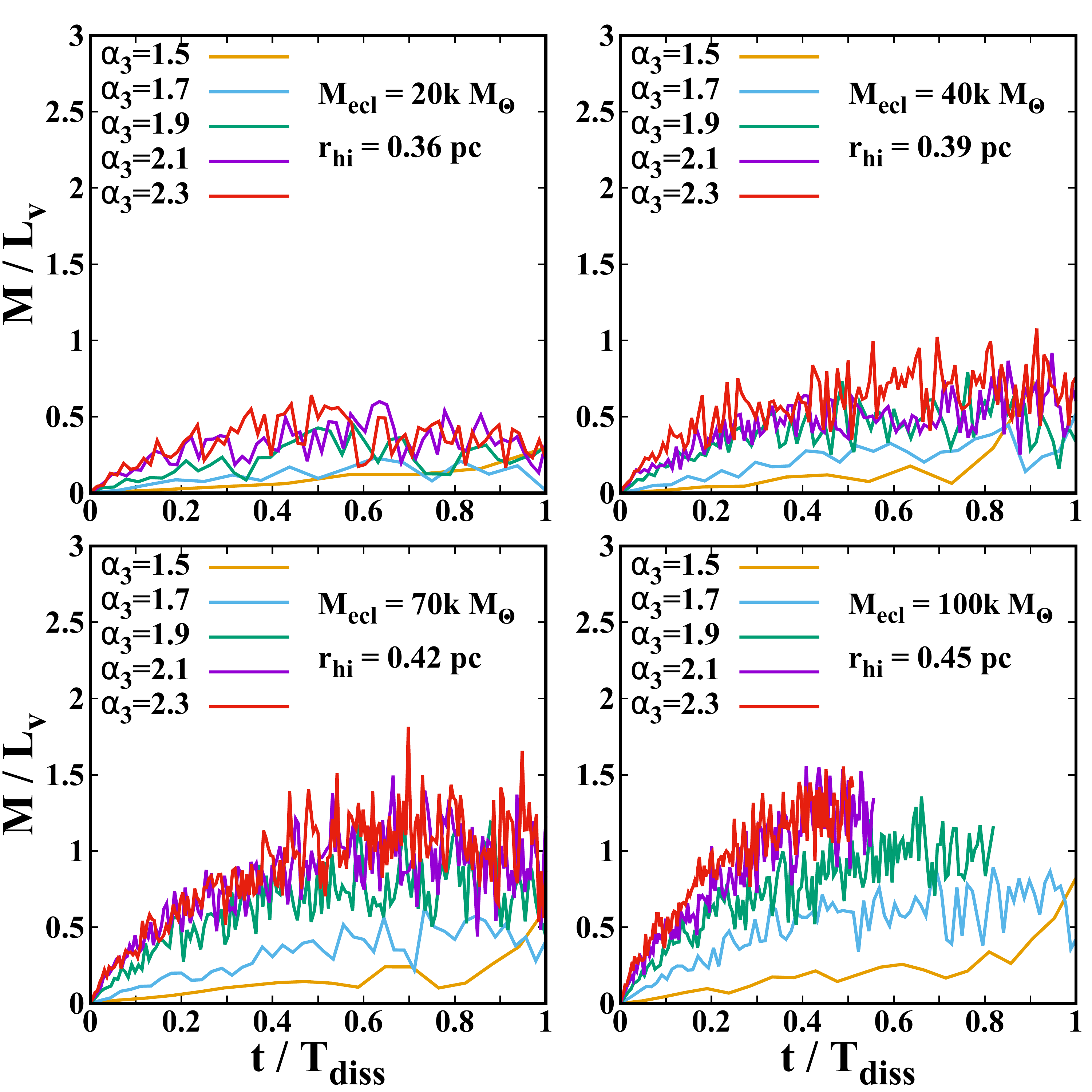}
		\caption{ The $M/L_V$  ratio vs. time for selected models with representative initial masses of 20k, 40k, 70k and 100k $ M_\odot$.  Different colors correspond to different value of  $\alpha_3$. Since our simulations were carried out until 14 Gyr, for models that survive longer, the curves are truncated (bottom  right panel). The dissolution times for these models are estimated by extrapolation (Table \ref{tab:details-table2}). }
		\label{fig:M/L-fig10}
		\end{figure}

	\subsection{The concentration parameter and density profile}
	
	The cluster central concentration parameter, which is defined as the logarithmic ratio of tidal to core radius
	\begin{equation}
		c=\log_{10}[r_t/r_c],
		\label{equ:c}
		\end{equation}
	has been used in various ways to characterize the dynamical state of GCs. The top-heaviness of the  IMF has an impact on the $c$-parameter.  In Figure \ref{fig:rc},   we show the evolution of the core radius and the $c$-parameter as a function of time for only four representative selected models to preserve the clarity of the figure. The core radius expands rapidly within the first few megayears, and hence  $c$ falls within the first few megayears of the evolution.
	
	As already shown by \cite{Brinkmann2017}, the expansion of the inner Lagrangian radii (and also the core radius) after gas expulsion  slows down or even stops after a few megayears of evolution, while  the outer layers expand, resulting in a slowly falling (in the case of a top-heavy IMF) or nearly constant (in the case of a canonical IMF) value of $c$.  The core radius  contracts during the  core-collapse phase in  models with a steeper slope (i.e., larger $\alpha_3$), resulting in an increase of the $c-$parameter. 
	
	Clusters with a top-heavy IMF experience a strong core expansion compared to the models with a canonical IMF. Therefore, for a given initial mass, clusters with a more top-heavy IMF have a relatively low concentration. Indeed, due to the abundant population of remnants in the central part of the clusters in the case of a top-heavy IMF, the core radius  is larger compared to star clusters with a canonical IMF. The formation of a  central  compact remnant subcluster might be relevant for the finding of the central peak in dispersion velocity of some GCs. The relevance of a top-heavy IMF for the formation of supermassive BHs has been discussed by \citet{Kroupa_2020}.
	
	Figure \ref{fig:alpha-concentration} compares the present-day global MF slope (in the mass range $0.3 -  0.8 M_{\odot}$ of the computed models, with the eyeball fit to the observed data of Galactic GCs  (shown as a red line) from  \cite{Demarchi2007} and \cite{Paust10}.   The canonical Kroupa IMF has $\alpha=1.72$ over this mass range. The computed models agree well with the general trend that less concentrated clusters typically show a shallower MF.   
	
	Figure \ref{fig:density-fig9} shows the mass density profile as a function of distance from the center of a cluster for two different snapshots (0 and 100 Myr). All clusters start with a Plummer profile. For a given initial cluster mass, the initial profile is independent of the IMF. As the clusters evolve,  the density profiles of the clusters become distinct for different $\alpha_3$ values such that for a  more top-heavy IMF the cluster reaches to a  lower density. As can be seen at 100 Myr, the number density in models with a canonical  IMF is higher than in those with a top-heavy IMF. The higher number density affects critically the two-body and three-body encounter rate, as well as the number of binaries. Models with the canonical IMF have a  higher fraction of dynamically formed binaries than top-heavy models.

	\begin{figure*}[]
		\centering
		\includegraphics[width=14cm]{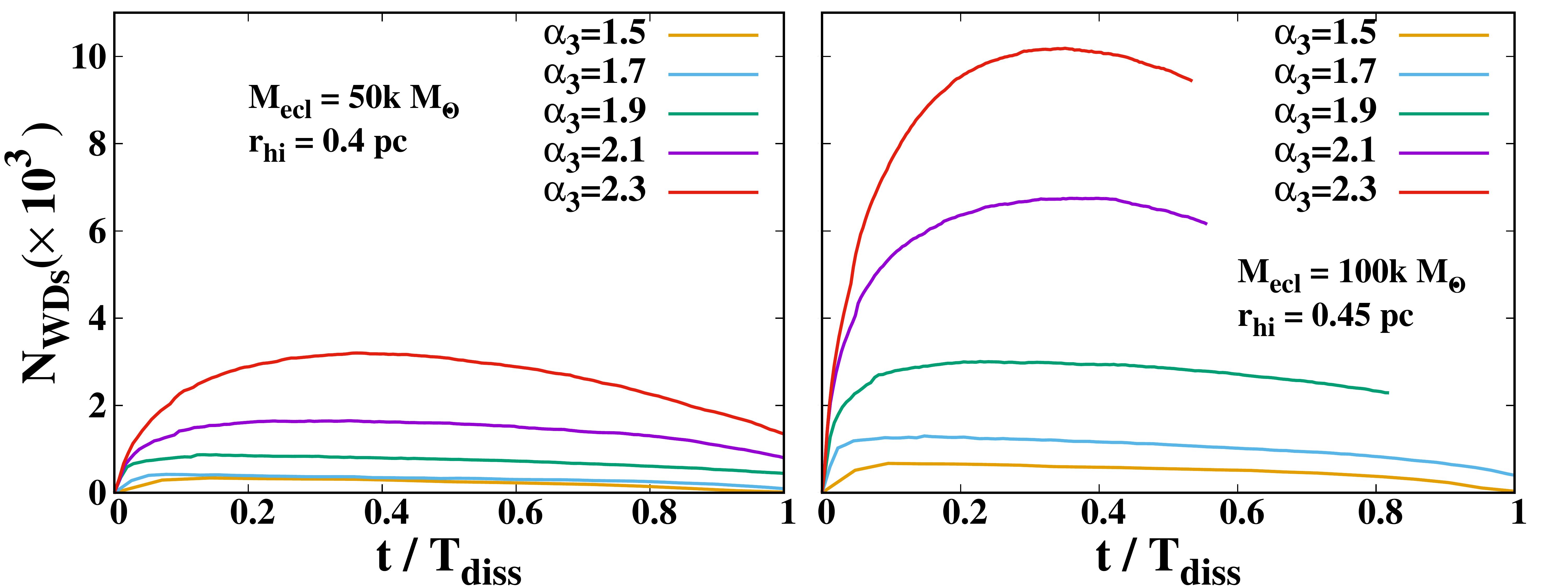}
		\includegraphics[width=14cm]{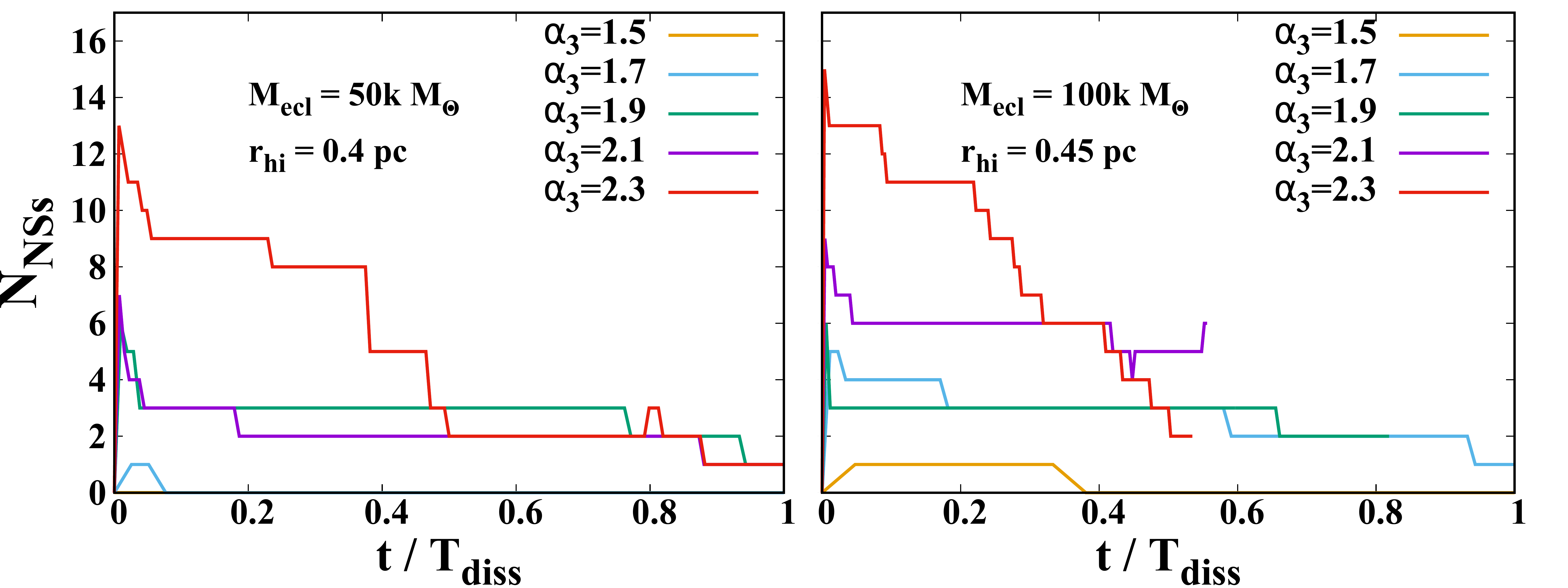}
		\includegraphics[width=14cm]{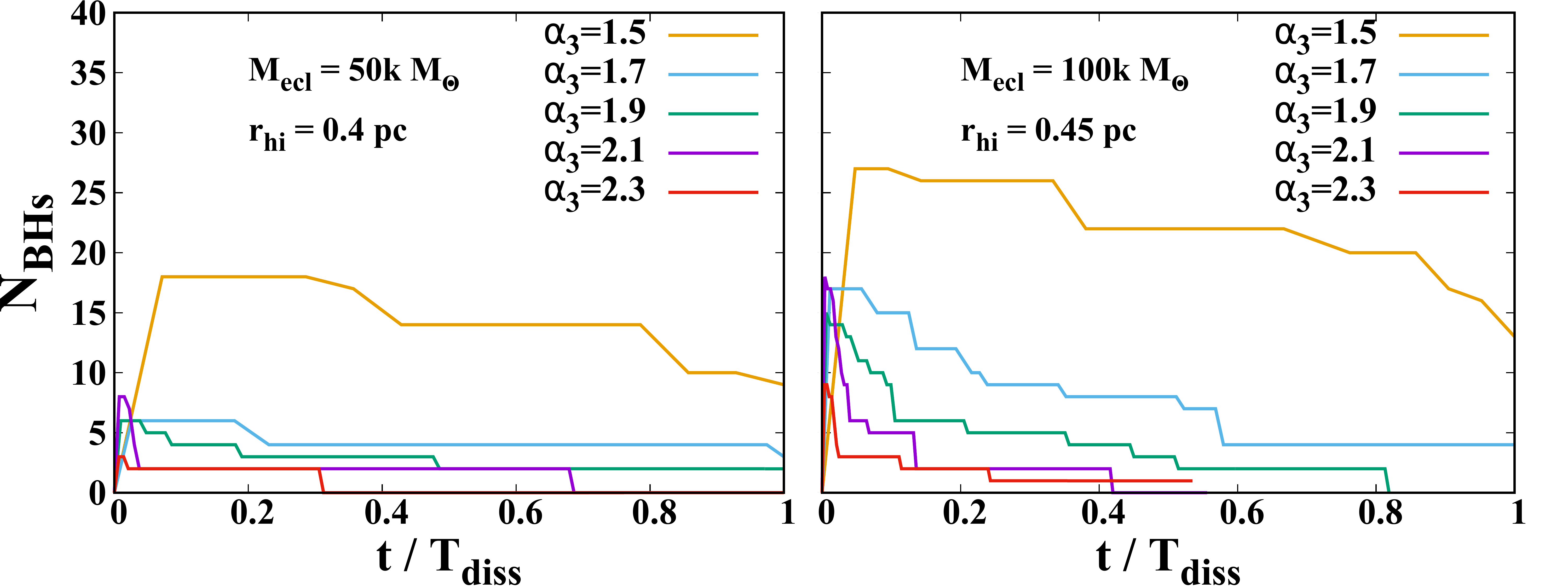}
		\caption{ Evolution of the number of bound WDs, NSs, and BHs as a function of time for clusters with different initial masses and $\alpha_3$. The color indicates the value for $\alpha_3$. Since our simulations were carried out until 14 Gyr, for models that survive longer, the curves are truncated (right panels). The dissolution time for these models is estimated by extrapolation (Table \ref{tab:details-table2}).}
		\label{fig:WDs-fig12}
		\end{figure*}

	\begin{figure*}[]
		\centering
		\includegraphics[width=8.5cm]{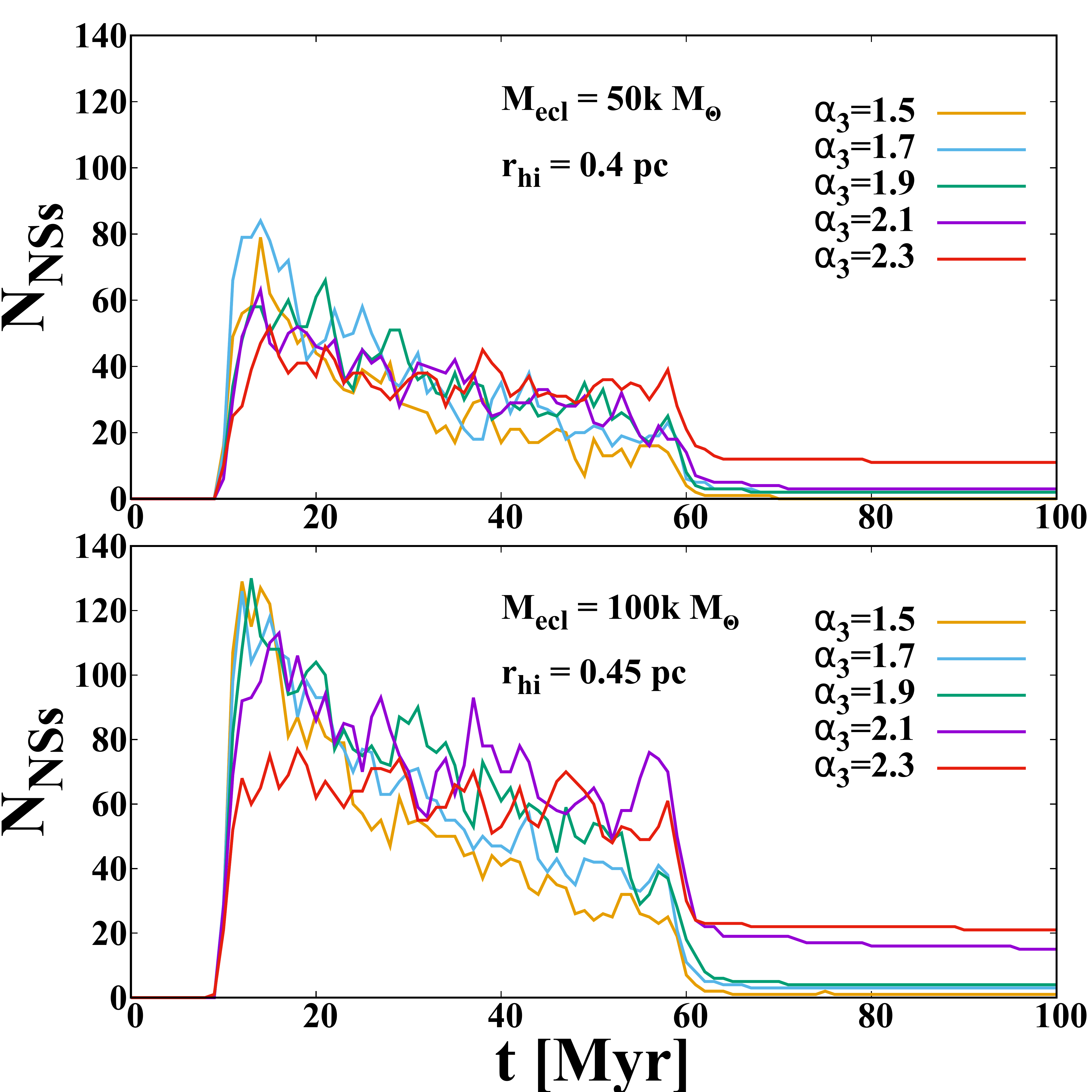}
		\includegraphics[width=8.5cm]{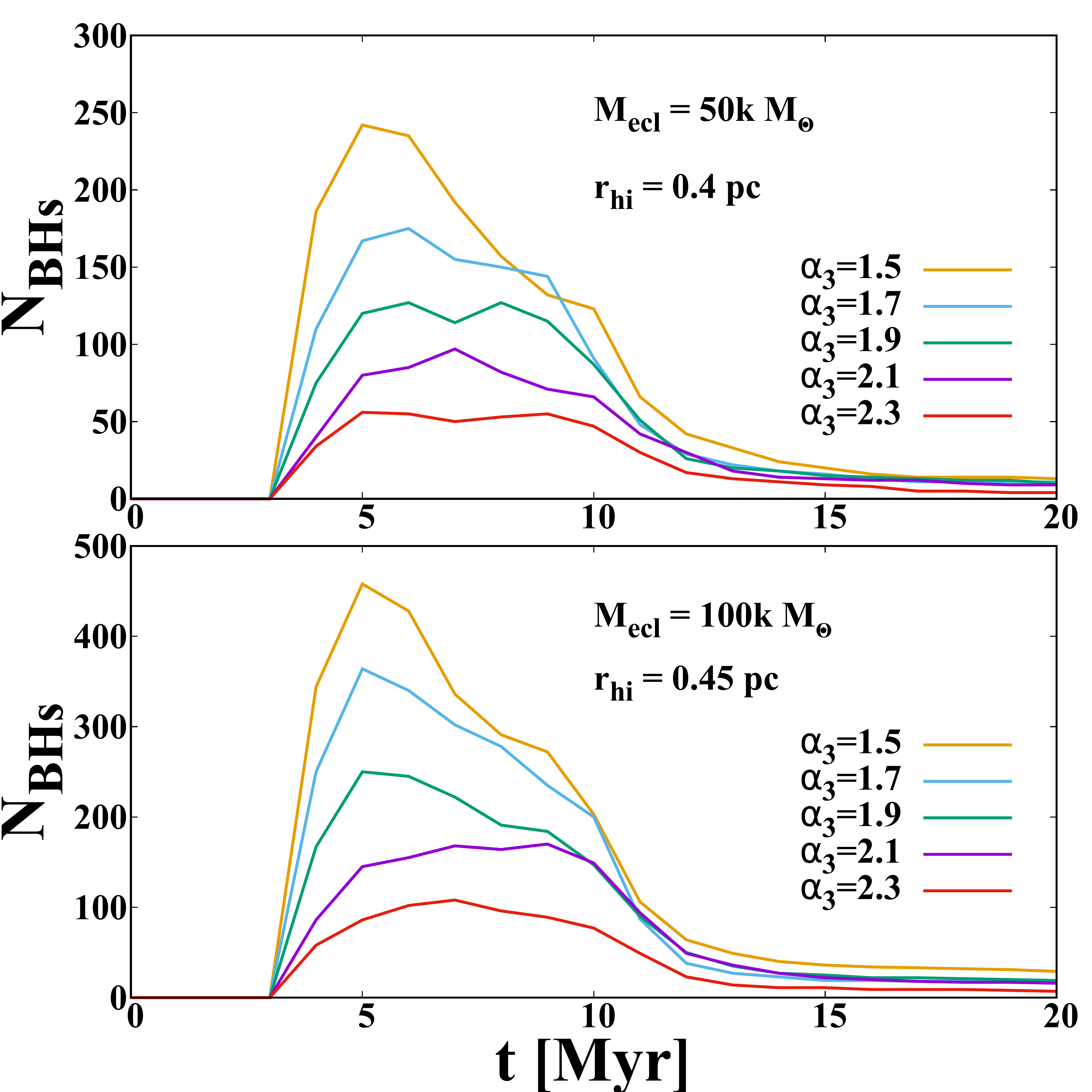}
		\caption{Early evolution of the total number of retained NSs and BHs as a function of time within the first 100 Myr for two models with different initial masses of 50,000 and 100,000 $M_{\odot}$ and  for various $\alpha_3$.}
		\label{fig:NS-2-t-100Myr}
		\end{figure*}

	\subsection{The mass-to-light ratio}
	
	%The mass fraction of low-mass stars increases in the beginning due to the mass loss of high-mass stars. Later their fraction decreases as aresult of the preferential depletion of low-mass stars.
	
	The present-day dynamical stellar mass-to-light ratio in the V-band ($M/L_V$) links the total stellar V-band luminosity of a system to its total mass and provides direct information about the present-day stellar population of a GC,  its total gravitational potential, and the IMF.  Stellar evolution has the long-term effect of increasing the $M/L_V$ since the massive stars gradually evolve into stellar remnants,  and hence the fraction of  high-mass  to low-mass stars decreases with time.  As a result of dynamical friction, the remnants that form through stellar evolution of the most massive stars  segregate into the cluster core. This leads to the cores of the older clusters being dominated by compact remnants.  On the opposite side, the  dynamical evolution  leads to the  preferred  escape of low mass stars from the star clusters \citep{VesperiniHeggie1997, BaumgardtMakino2003, Baumgardt17}. However, \cite{Bianchini17} showed that the decrease of $M/L_V$ is primarily driven by the dynamical ejection of dark remnants, rather than by the escape of low-mass stars.  %Both these effects (mass segregation and preferential loss of low-mass stars) have direct implications on shaping the M/L value of GCs. 

	Figure \ref{fig:M/L-fig10} depicts the evolution of the $M/L_V$ ratios of star clusters as a function of time $(t/T_{diss})$  for different initial cluster masses and different   values of $\alpha_3$. $M$ contains all stars and remnants bound to the cluster. In general, all $M/L_V$ ratios are increasing, although with considerable differences between individual models. Differences arise since clusters that dissolve quicker do so at an age in which they still contain many luminous stars.

	For massive clusters that last longer (see e.g., clusters with $M_{ecl}$ = 40k and 70k $M_{\odot}$ in Figure \ref{fig:M/L-fig10}), the dynamical evolution leads to a decrease of the average  $M/L_V$ ratio by a value of about 0.5 at $t/T_{diss}=0.8$.  As a cluster ages,  the fraction of compact remnants increases in the case of $\alpha_3=1.5$, so that, close to the end, the $M/L_V$ ratio increases again.

	\subsection{The number of remnants} \label{sec:remnants}
	
	In the early stage of the cluster evolution the number of dark remnants (WDs, NSs, and BHs) increases as stars evolve.  The number of dark remnants  remaining in a cluster depends primarily  on the density of the cluster, which is directly related to the frequency of gravitational encounters that the dark remnants experience \citep{Hurley16}. The different densities  that clusters with different  $\alpha_3$  reach owing to the  different expansion rate also have an impact on the final number of retained BHs and NSs (see Figure \ref{fig:density-fig9}).

	It is well known that the presence of BHs and NSs can affect the overall structure and evolution of the parent cluster \citep{Contenta2015}. In order to see how the number of compact remnants changes in our clusters, we plot the evolution of the number of WDs, NSs, and BHs as a function of cluster age (Figure \ref{fig:WDs-fig12}). 
	
	In this work we consider a Maxwellian dispersion of kick velocities, $\sigma_{BH}=190~$km s$^{-1}$, which means that almost all BHs and NSs are kicked out of the clusters  \citep{HansenPhinney, Jerabkova_2017, Pavlik_2018}.  Moreover, we calculate one model with $\sigma_{BH}=0$ km s$^{-1}$ for comparison.

	\subsubsection{WDs}
	The number of WDs in a star cluster rises in the first part of the cluster's life as it evolves and then declines in the later part owing to their dynamical ejection. It has been shown that more than half of the total cluster mass can be in the form of WDs close to the end of its lifetime, depending on the IMF and $T_{diss}$ \citep{Giersz01}. Figure \ref{fig:WDs-fig12} shows  the overall evolution of the number of WDs as a function of time. The number of WDs is smaller in the clusters with an initially top-heavy IMF.  Indeed, in a cluster with an initially top-heavy IMF the dissolution time is shorter and a smaller fraction of stars are turned into WDs as a result of stellar evolution. But in a cluster with a canonical IMF and thus a longer dissolution time, say, a cluster with $M_{ecl}=100 K M_{\odot}$, WDs become the dominant mass group after roughly 80$\%$ of the cluster stars are lost. In all panels with different initial mass, the number of WDs in  models with a canonical IMF  is higher than in models with a top-heavy IMF. The large population of WDs  that contribute in mass but not in light results in a larger dynamical  $M/L_V$ ratio of clusters that were born with a  less top-heavy IMF,  as already shown in Figure \ref{fig:M/L-fig10}. 
	
	\subsubsection{NSs}
	%As can be seen in Figure \ref{fig:black-hole-n/n0-fig12} we show the fraction of BHs over the initial number of stars as a function of dissolution time.
	In order to shed light on the early evolution of the NS population,  the number of bound NSs is plotted in Figure \ref{fig:WDs-fig12}  as a function of time within the first 100 Myr. As can be seen, the production of the NSs begins at about 10 Myr and ends at about 60 Myr.  In all panels, the number of NSs in a top-heavy model is higher than in canonical IMF models during the first 20 Myr.  But, due to the faster evolution of heavy stars and higher expansion rates in top-heavy models,  most of them will be unbound.  Therefore, ultimately, the number of retained NSs is higher in models with a canonical IMF.
	Moreover, in Figure \ref{fig:WDs-fig12}, we plot the evolution of the total number of bound NSs until the dissolution time for different models.

	%Less relaxed clusters have a higher fraction of dark remnants than more relaxed clusters.
	\subsubsection{BHs}
	Stellar mass BHs form in the cluster within the first 10 Myr. 
	The evolution of the total number of BHs retained  in the clusters is shown in Figure \ref{fig:WDs-fig12}. The number of BHs grows to the maximum value in the first 5 Myr, when BHs begin to form, and is then followed by a slow decline over time until there are just a few (right panels of Figure \ref{fig:NS-2-t-100Myr}). The number of BHs begins to decrease all the way to 12 Gyr as they are ejected through strong binary encounters in the core.  The majority of our models with a top-heavy IMF ($\alpha_3=1.5$) end with roughly $5-15$ BHs. Higher-$N$ models have more BHs at the end of their lifetime than lower-$N$ models. They retain about  $50 \%- 90\%$  of their initial  BHs.  According to Figure \ref{fig:WDs-fig12}, models with a larger $\alpha_3$ retain  fewer  BHs  at  12 Gyr. Almost all BHs are ejected from the clusters with $\alpha_3  \geq 1.9$ before dissolution. 
	
	As can be seen in Figure \ref{fig:WDs-fig12}, the number of BHs declines faster in clusters with a  higher density (e.g., models with $M=100k M_{\odot}$).  This is because,  while a cluster reaches a more advanced state of dynamical evolution, the number of stellar encounters that the dark remnants have experienced is higher, and therefore they have experienced a higher number of dynamical ejections.

	\subsubsection{The effect of the retention fraction}
	
\begin{figure}[]
		\centering
		\includegraphics[width=8.2cm]{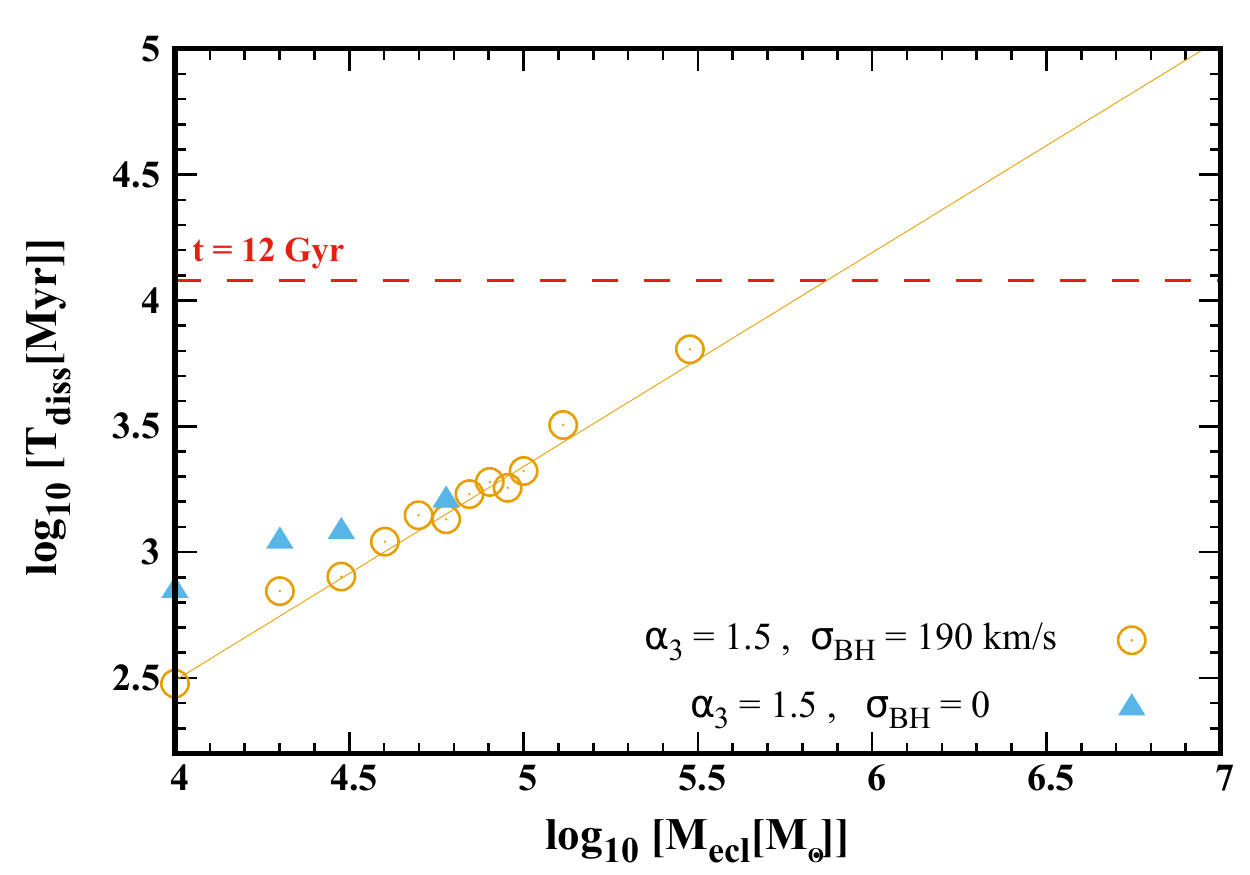}
		\caption{ Dissolution time as a function of the initial cluster mass for two different values of $\sigma_{BH} = 190~km/s$ (yellow open circles) and $\sigma_{BH} = 0 $ (blue filled triangles). For all models, a top-heavy IMF with $\alpha_3=1.5$ is considered.  }
		\label{fig:dissolution-BHs-fig14}
		\end{figure}

	These results, however, depend on the BH retention fraction. In order to assess how sensitive the results are on the BH retention fraction, we recalculate some models with the same initial conditions but assuming zero kick velocity,  $\sigma_{BH}=0$ (i.e., 100\% retention fraction for the BHs). Figure \ref{fig:dissolution-BHs-fig14} depicts the cluster lifetimes, $T_{diss}$, as a function of the initial mass of the star clusters for $\alpha_3 = 1.5$. As can be seen, the dissolution time of clusters with $M_{ecl} < 10^{4.8} M_{\odot}$ is longer for clusters with a higher retention fraction of BHs. 
	
\begin{figure*}[]
		\centering
		\includegraphics[width=14.3cm]{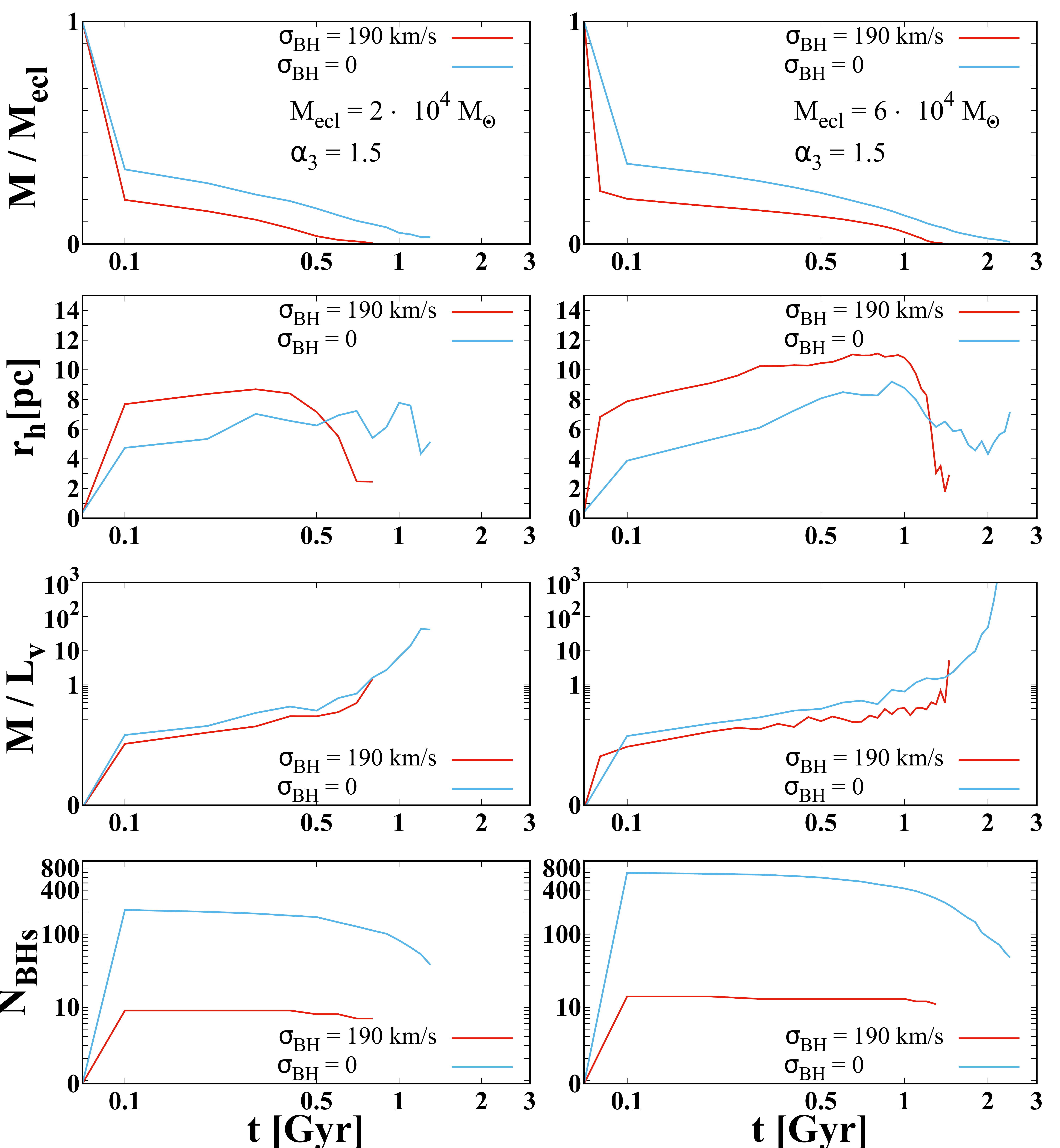}
		\caption{ Comparison of the evolution of the half-mass radius, the total mass of cluster, the $M/L_V$ ratio and the number of BHs for $\sigma_{BH} = 190~km/s$ (red line) and $\sigma_{BH} = 0 $ (blue line) for two clusters with initial masses of 20k and 60k $M_{\odot}$. The IMF has $\alpha_3 = 1.5$ for both clusters.}
		\label{fig:4plot-BHs-different-RF-fig13}
		\end{figure*}

	Figure \ref{fig:4plot-BHs-different-RF-fig13} compares the half-mass radius, the total mass of the cluster, the mass-to-light ratio ($M/L_V$), and the number of BHs as a function of time for two star clusters with $\alpha_3 = 1.5$ and two different initial retention fractions of BHs. As can be seen, a cluster with a 100\% BH retention fraction will collect more BHs and reach a higher value of $M/L_V$ compared to a model with zero retention fraction. Retention fractions  dependent on $M_{ecl}$ are discussed further in \cite{Jerabkova_2017} and  \cite{Pavlik18}.

	\subsection{The applicability of the results}

	Since the top-heaviness of the IMF depends on the metallicity and density (metal-poorer environments appear to form flatter IMFs), the present-day characteristics of the metal-poor massive GCs (after the long-time stellar and dynamical evolution) might be better explained by assuming a top-heavy IMF at birth. In this case, the more massive GCs form relatively more massive stars.  To calculate the present-day characteristics of massive metal-pour GCs, the long-term dynamical evolution should be considered based on the results of our modeled clusters  with top-heavy IMFs. Until now the only available formalism for cluster lifetimes was the one developed by \citet{BaumgardtMakino2003} in which clusters with a canonical IMF and without early gas expulsion have been considered.
	
	In order to match any particular cluster or population of clusters, the present model results need to be applied to the specific properties of the targets. Thus, for example, Segue~3 is regarded as the youngest GC in the outer halo with an estimated age of 3.2 Gyr \citep{Ortolani_2013}. Thus, the appropriate metallicity and age will need to be selected from the present model grid in order to compare the models with Segue~3. The inferred required tidal boundary conditions to match Segue~3 at an age of~3.2~Gyr will then constrain the environment in which the cluster may have been born. As another example, the cluster Whiting~1 that is associated with the Sgr dSph has an age of 6.5 Gyr \citep{Carraro_2007}, and so the above will need to be repeated for this particular cluster to see whether it can be matched by a model from the grid. Further, a recent analysis of the population of galactic GCs found that 35 out of 151 objects have disk kinematics \citep{Bajkova_2020}, and the disk GCs will need to be compared with that part of the grid that contains the appropriate tidal boundary conditions. Also, note that the GC population  contains the sub-category ``young halo GC" (YHGC). These are more metal-poor but younger than the Population~II halo GCs,  follow the same spatial and kinematic distribution as the dwarf galaxies  in the disk of satellites \citep{Pawlowski_2012} and may have formed in the tidal debris during a past encounter between the Milky Way and Andromeda \citep{Bilek_2018} .
	
	%Indeed, finding observational evidence for a varying high-mass IMF is difficult as stars more massive than one solar mass have long since evolved away from the main sequence and cannot be observed. In order to describe the PD characteristics of some Galactic GCs, top-heavy IMFs need to be invoked (Marks & Kroupa 2010). 
	
	The large mass-to-light ratios in some ultracompact dwarf galaxies (UCDs)  indicate that these  formed with a top-heavy IMF as well \citep{Dabringhausen_2008}. Given the ideas of the formation of UCDs from star cluster complexes \citep{Fellhauer_2005} or as the most massive GCs \citep{Mieske_2002}, the present-day characteristics of UCDs  can be explained by invoking top-heavy IMFs. The simulations reported here can be useful to calculate the early expansion of UCDs due to the evolution of massive stars in the first 100 Myr (Mahani et al., in preparation).

	%In the framework of the theory of the integrated galactic IMF (IGIMF, i.e. the stellar IMF of whole galaxies; Kroupa & Weidner 2003; Weidner & Kroupa 2005), a similar parametrization of systematically varying IMF has been used (Weidner, Kroupa & Pflamm-Altenburg 2011). The IGIMF, being the sum of all IMFs in all young star clusters, then also becomes top heavy. 

	%%%%%%%%%%%%%%%%%%%%%%%%%%%%%%%%%%%%%%%%%%%%%
	\section{CONCLUSIONS} \label{sec:conclusions}
	
	Several independent pieces of observational and theoretical evidence have been put forward in the past decade, showing that the stellar IMF may depend on the star formation environment such that it becomes increasingly top-heavy with decreasing metallicity and increasing gas density of the forming object. This can significantly affect the evolution of star clusters and their survival/disruption. The previous modeling of GCs usually assumes a universal canonical IMF. Using the state-of-the-art \textsc{nbody6} code,  we have performed a large grid of simulations of multimass star clusters moving through the host galaxy to study the impact of varying the degree of top-heaviness on the survival rate and final properties of star clusters. The mass loss due to the stellar evolution, the two-body relaxation, the residual gas expulsion,  and  the tidal field of the host galaxy are included in our simulations. Our results can be summarized as follows:
	
	\begin{itemize}
		
		\item All simulated clusters dissolve mainly as a result of two-body relaxation-driven evaporation and tidal truncation, with the dissolution time of clusters  scaling proportionally to the embedded cluster mass to a power $x\approx 0.85$ which is in agreement with the results of \citet{BaumgardtMakino2003} for the scaling of the lifetimes of clusters in tidal fields. The scaling law does not change significantly if one goes from the models with a canonical IMF to those with a top-heavy IMF. 
		
		\item Both the initial embedded cluster mass  and the degree of top-heaviness have a strong influence on the evolution of star clusters. All star clusters expand owing to the early gas expulsion that leads to an unbinding of stars from the star clusters. For various $\alpha_3$, star clusters expand by a factor of 15 - 30 in the case of the canonical to most  top-heavy IMF considered here, respectively. We also showed that the details of this process  depend on the starting conditions, such as maximum stellar mass and BH retention fraction.
		
		\item  Utilizing the mass-radius relation of \citet{MarksKroupa2012}  for initializing the models as embedded clusters, and by varying the degree of top-heaviness, we calculated the minimum cluster mass needed for the cluster to survive at least 12 Gyr of evolution. 
		
		\item  We showed that the cluster final half-mass radii, bound mass fraction, and their survival rates are strongly influenced by the degree of top-heaviness. Decreasing the slope of the IMF at the high-mass end (making the clusters more top-heavy) leads to a faster dissolution of the star clusters. We additionally find the best-fitting equation for $T_{diss} - M_{ecl}$ for different $\alpha_3$. %\textbf{TB-Done: Utilizing our results for the lifetimes, we predict that a significant fraction of all Galactic star clusters might have been destroyed within the Hubble time.}
		
		\item  The mass and number fraction of WDs, NSs,  and BHs increase throughout cluster evolution since these \textbf{objects} are the most massive components and are therefore least likely to escape. For GCs near the end of their lifetime,  most of the cluster mass should be in the form of compact remnants.  Models with a top-heavy IMF create and collect more BHs, while  more WDs are created and finally retained in the cases with a canonical IMF. As a result of the removal of stars from the outer parts of a cluster by the tidal field of the host galaxy and spiraling of stellar-mass BHs toward the center of their clusters due to dynamical friction, the BHs that formed through stellar evolution of the most massive stars segregate into the cluster's  core and form a subsystem of BHs (or ``dark star clusters," as named for the first time by \citealt{Banerjee_2011}). The formation of such a BH subsystem is already well studied in star clusters with a canonical IMF \citep{Banerjee_2011, Breen_2013} and is also of interest in the context of galactic nuclei \citep{Kroupa_2020}. In the future, we will explore the effect of the top-heaviness of the IMF, as well as different values of the compact remnant retention fraction, on the formation and the lifetime of dark star clusters.
		
		%("dark clusters", \citealt{Banerjee_2011})
		
		\end{itemize}
	
	%The simulations reported here should be useful for a number of follow-up projects. First, to verify or test the analysis of the \cite{MarksKroupaDabringhausen2012} relation for  the $\alpha_3$ dependence on metallicity and density of the star-forming cloud. Our computations also allow us to study the impact of various $\alpha_3$ on the evolution of whole embedded cluster systems in galaxies: starting with the initial half-mass radius of the clusters by the \cite{MarksKroupa2012} relation and gas-expulsion, the mass fraction remaining bound to each individual cluster, the half-mass radius, the dissolution time and the expansion rate can be calculated by interpolating in our grid. Repeating the process for all clusters will then give the impact of initial mass and degree of top-heaviness on the whole star cluster system as a function of the age of the system.

	The simulations reported here should be useful for a number of follow-up projects. First, it will be important to test the analysis of the GC data used by \cite{MarksKroupaDabringhausen2012} to infer the relation for  the $\alpha_3$ dependence on metallicity and density of the star-forming cloud. The original analysis inferred the top-heaviness of the IMF from the observational data through the discovery of a correlation between the metallicity and inferred birth density with the value of $\alpha_3$. This was achieved by calculating the energy the massive stars needed to provide in order to expand the mass segregated cluster and unbind the low-mass stars from the cluster through gas expulsion. But it remains to be checked if such an IMF will actually, in $N$-body simulations, lead to the observed cluster properties. The results documented in the present contribution also allow us to study the impact of various $\alpha_3$ values on the evolution of whole embedded cluster systems in galaxies: starting with the initial half-mass radius of the clusters being given by the \cite{MarksKroupa2012} relation and gas-expulsion, the mass fraction in stars remaining bound to each individual cluster, the half-mass radius of each cluster, the dissolution time and the expansion of each cluster can be calculated by interpolating in our grid. By matching the initial distributions (in mass and radius) of the model cluster population to the observed present-day properties of the GCs it will be possible to infer the fraction of clusters that dissolved, which mass the initial population had and will allow a test of this whole scenario by comparing the model-implied population~II halo with the observed halo stellar population. This last point is important because the model will need to match the appropriate (population~II) observed GC population and also at the same time the halo population~II assuming (in this model) that the population~II halo has its origin in the dissolved embedded clusters.
	
	The top-heavy IMF will produce more BH binaries, but this  is not discussed here and will be studied as an upcoming project. The primordial mass segregation also plays an important role in the early and long-time evolution of star clusters, especially if they are set up with a top-heavy IMF.

%% For this sample we use BibTeX plus aasjournals.bst to generate the
%% the bibliography. The sample63.bib file was populated from ADS. To
%% get the citations to show in the compiled file do the following:
%%
%% pdflatex sample63.tex
%% bibtext sample63
%% pdflatex sample63.tex
%% pdflatex sample63.tex

\bibliography{apjReferences}{}
\bibliographystyle{aasjournal}

\appendix \section{Table of results}
In Table \ref{tab:details-table2}, we summarize the initial cluster properties and the key results of the simulated clusters.

\startlongtable
\begin{deluxetable*}{ccccccccccccc}
	\tablecaption{Details of the $N$-body Computations That Lead to the Formation of a Bound Cluster on a Circular Orbit. The first column gives $\alpha_3$. The next columns give, respectively, the initial stellar mass of the models ($M_{ecl}$), the initial half-mass radius, the initial number of stars  per cluster, the initial crossing time, the tidal radius,  the ratio of $r_{hi}/r_t$, the value of the dissolution time in Gyr and the initial half-mass relaxation time in Myr. Columns 10 to 13 contain only the clusters surviving for  longer than 12 Gyr and give the ratio of the half-mass radius at 12 Gyr to the initial half-mass radius, the ratio of the mass of the cluster at 12 Gyr to the initial stellar mass of the cluster, the value of the slope of the mass function $\alpha$ for the stars with masses between $0.2 M_{\odot}$ until $0.8 M_{\odot}$ at 12 Gyr ($\alpha_3= +2.35$ would be the Salpeter index), and column 13 gives the half-mass relaxation time of the surviving clusters at 12 Gyr (columns 10 to 13 are empty for clusters that dissolve within 12 Gyr). \tablenotemark{} \label{tab:details-table2}}
	\tablehead{
		\colhead{$\alpha_{3}$} &\colhead{$M_{ecl}$} &\colhead{$r_{hi}$} &\colhead{$N$} &\colhead{$t\sb{c}$} &\colhead{$r_t$} & \colhead{$r_{hi}/r_t$} & \colhead{$T\sb{diss}$} & \colhead{$t\sb{rh}^{0}$} & \colhead{$r\sb{hf}  / r\sb{hi}  $} & \colhead{$M\sb{f}  / M\sb{ecl} $} &
		\colhead{$\alpha  $} & \colhead{$t\sb{rh}^{12}$} \\
		\colhead{} &\colhead{$[ 10^3 M_{\odot}]$} &\colhead{$[pc]$} &\colhead{} &\colhead{$[Myr]$} &\colhead{$[pc]$} &\colhead{$\times 10^{-3}$} &\colhead{$[Gyr]$}& \colhead{$[Myr]$} &\colhead{}&
		\colhead{} &
		\colhead{$[0.2<\frac{m}{M_{\odot}}<0.8]$} &
		\colhead{$[Gyr]$}
	}
	\colnumbers
	\startdata
	 1.5  &   10 &  0.33 &  4379  &  0.058  &  31.7  &  11  &  0.3  &  1.4  &  -&-&-&-\\
	 1.7  &   10 &  0.33 & 7646  &  0.058  &  31.7  &  11  &  0.7  &  2.5 &  -&-&-&-\\
	 1.9  &   10 &  0.33 &  10473  &  0.058  &  31.7  &  11  &  2.0  & 3.4  &  -&-&-&-\\
	 2.1  &   10 &  0.33 &  14869  &  0.058  &  31.7  &  11  &  3.3  &  4.9 &  -&-&-&- \\
	 2.3  &   10 &  0.33 & 17337  &  0.058  &  31.7  &  11  &  4.3  & 5.6 &  -&-&-&-\\
	%%%%%
	 1.5  &   20 &  0.36 &  9857  &  0.047  &  40.0  &  9.1  &  0.7  &  2.6 &  -&-&-&-\\
	 1.7  &   20 &  0.36 &  14832  &  0.047  &  40.0  &  9.1  &  1.6  & 3.9 &  -&-&-&-\\
	 1.9  &   20 &  0.36 & 21754  &  0.047  &  40.0  &  9.1  &  3.8  &  5.7 &  -&-&-&-\\
	 2.1  &   20 &  0.36 & 29713  &  0.047  &  40.0  &  9.1  &  6.3  &  7.8 &  -&-&-&-\\
	 2.3  &   20 &  0.36 &  38132  &  0.047  &  40.0  &  9.1  &  6.8  &  10.0 &-&-&-&-\\
	%%%%%
	 1.5  &   30 &  0.38 & 13970  &  0.042  &  46.0  &  8.4  &  0.8   &  3.2 &-&-&-&-\\
	 1.7  &   30 &  0.38 &  21465  &  0.042  &  46.0  &  8.4  &  2.7  & 5.0 &-&-&-&-\\
	 1.9  &   30 &  0.38 & 34072  &  0.042  &  46.0  &  8.4  &  7.0 & 8.0 &-&-&-&-\\
	 2.1  &   30 &  0.38 & 44247  &  0.042  &  46.0  &  8.4  &  9.1  & 10.3 &-&-&-&-\\
	 2.3  &   30 &  0.38 & 55594  &  0.042  &  46.0  &  8.4  &  9.5 & 12.9 &-&-&-&-\\
	%%%%%%%%
	 1.5  &   40 &  0.39 & 18726  &  0.038  &  50.0  &  7.9  &  1.1  & 3.9 &-&-&-&-\\
	 1.7  &   40 &  0.39 & 30297  &  0.038  &  50.0  &  7.9  &  2.7 & 6.3 &-&-&-&-\\
	 1.9  &  40 &  0.39 & 43493  &  0.038  &  50.0  &  7.9  &  8.0  & 9.0 &-&-&-&-\\
	 2.1  &   40 &  0.39 & 58750  & 0.038  &  50.0  &  7.9  & 11.4 & 12.3  &  7.9 & 0.03 & -1.13$\pm$ 0.11 & 0.051 \\
	 2.3  &   40 &  0.39 &  72728  & 0.038  &  50.0  &  7.9  & 12.8 & 15.2 &  9.6 & 0.07 & -0.23$\pm$ 0.07 & 0.130 \\
	%%%%%
	%%%%%%%%
	 1.5  &   50 &  0.40 & 24589  &  0.035  &  54.0  &  7.6  &  1.4   & 4.8 &-&-&-&-\\
	 1.7  &   50 &  0.40 & 36889  &  0.035  &  54.0  &  7.6  &  3.9  &  7.2 &-&-&-&-\\
	 1.9  &  50 &  0.40 & 57012  &  0.035  &  54.0  &  7.6  &  10.5  & 11.1 &  7.3 & 0.005 & -1.52$\pm$ 0.36 & 0.017  \\
	 2.1  &   50 &  0.40 &  73043  &  0.035  &  54.0  &  7.6  &  13.4 & 14.2 &  10.6 & 0.07  & 0.0$\pm$ 0.01 & 0.185 \\
	 2.3  &   50 &  0.40 &  91113  &  0.035  &  54.0  &  7.6  &  14.4  & 17.7 &  11.7 & 0.01  & +0.3$\pm$ 0.05 & 0.280 \\
	%%%%%
	1.5 &  60 & 0.41 &28281 &0.033 &57.0 & 7.3 & 1.4 &5.2&-&-&-&-\\
	1.7 & 60 & 0.41 &44614 &0.033 &57.0 & 7.3 & 4.7 &8.2&-&-&-&-\\
	1.9 & 60 & 0.41 &66636 &0.033 &57.0 & 7.3 & 11.7 &12.3& 8.9&0.03&-0.42 $\pm$ 0.08 &0.081\\
	2.1 & 60 & 0.41 &79579 &0.033 &57.0 & 7.3 & 16.0& 14.6& 12.4&0.12&+0.51 $\pm$ 0.07&0.386 \\
	2.3 & 60 & 0.41 &111158 &0.033 &57.0 & 7.3 & 17.1&20.4& 13.7&0.18&+0.75$\pm$ 0.07&0.593  \\
	%%%%%
	%%%%%
	1.5 & 70& 0.42 &32323 & 0.032 & 60.0 & 7.1& 1.7&5.7&-&-&-&-\\
	1.7 & 70& 0.42 &51366 & 0.032 & 60.0 & 7.1&4.5&9.1&-&-&-&-\\
	1.9 & 70& 0.42 &78889 & 0.032 & 60.0 & 7.1&13.2&13.9& 12.1&0.05&+0.18 $\pm$ 0.03&0.235 \\
	2.1 & 70& 0.42 &102966 & 0.032 & 60.0 & 7.1&17.3 &18.2& 13.4&0.16&+0.80 $\pm$ 0.08&0.619\\
	2.3 & 70& 0.42 &126627 & 0.032 & 60.0 & 7.1&19.2&22.3& 14.0&0.22& +0.89 $\pm$ 0.05&0.784 \\
	%%%%%%%%
	%%%%%
	1.5 & 80& 0.43 & 37074 & 0.03 & 63.0&6.9 &1.9 &6.3&-&-&-&- \\
	1.7 & 80& 0.43 &58745 & 0.03 & 63.0&6.9 &6.0 &10.0&-&-&-&- \\
	1.9 & 80& 0.43 &86008 & 0.03 & 63.0&6.9 &14.8 &14.7& 12.0&0.07&+0.46 $\pm$ 0.06&0.337 \\
	2.1 & 80& 0.43 &119240 & 0.03 & 63.0&6.9 &20.0&20.4& 14.0&0.18&+0.96 $\pm$ 0.03&0.809 \\
	2.3 & 80& 0.43 &147388 &0.03 & 63.0&6.9 &23.0 &25.2& 14.1&0.2&+1.21 $\pm$ 0.05&1.128 \\
	%%%%%%%%
	%%%%%%%%
	1.5 & 90& 0.44 & 41750 & 0.029 & 65.0 &6.7&1.8&6.7 &-&-&-&-\\
	1.7 & 90& 0.44 & 65973 & 0.029 & 65.0 &6.7& 6.2&11.0&-&-&-&-\\
	1.9 & 90& 0.44& 100596 & 0.029 & 65.0 &6.7&16.3 &16.8& 12.4&0.09&+0.75 $\pm$ 0.06&0.497 \\
	2.1 & 90& 0.44 & 132773 & 0.029 & 65.0 &6.7&22.4&22.2& 14.2&0.21&+1.06 $\pm$ 0.08&0.993\\
	2.3 & 90& 0.44 & 165458 & 0.029 & 65.0 &6.7&26.3&27.6& 14.4&0.27&+1.14 $\pm$ 0.08&1.214\\
	%%%%%
	%%%%%%%%
	1.5 & 100& 0.45 & 46328 & 0.028 & 68.0 &6.5 & 2.1 &7.6&-&-&-&-\\
	1.7 & 100& 0.45 & 75236 & 0.028 & 68.0 &6.5 & 8.8 & 12.3&-&-&-&-\\
	1.9 & 100& 0.45 & 110149 & 0.028& 68.0 &6.5 & 17.1 &18.0& 14.2&0.11&+0.88 $\pm$ 0.08&0.744\\
	2.1 & 100& 0.45 & 146424 & 0.028 & 68.0 &6.5 & 25.0&24.0&14.4&0.23&+1.13 $\pm$ 0.09&1.200\\
	2.3 & 100& 0.45 & 182453 & 0.028 & 68.0 &6.5 & 28.8&30.0& 13.2&0.31&+1.17 $\pm$ 0.08&1.238\\
	%%%%%
	1.5 & 130& 0.46 & 60783 & 0.026 & 74.0 &6.2 & 3.2 &9.0&-&-&-&-\\
	1.5 & 300& 0.52 & 140982 & 0.018 & 97.0 &5.3 & 6.4 &16.5&-&-&-&-\\
	%%%%%%%
	\enddata
	%\tablenotetext{a}{Adjusted for inflation}
	%\tablenotetext{b}{Accounts for the change from page charges to digital quanta in April, 2011}
	%\tablecomments{Note that {\tt \string \colnumbers} does not work with the
	%vertical line alignment token. If you want vertical lines in the headers you
	%can not use this command at this time.}
	\end{deluxetable*}

\appendix \section{GPU run time}

Direct $N$-body simulations of  the evolution of dense star clusters with small initial half-mass radii are  very time- consuming even on GPU-based supercomputers.

\begin{figure}[h]
	\centering
	\includegraphics[width=85mm]{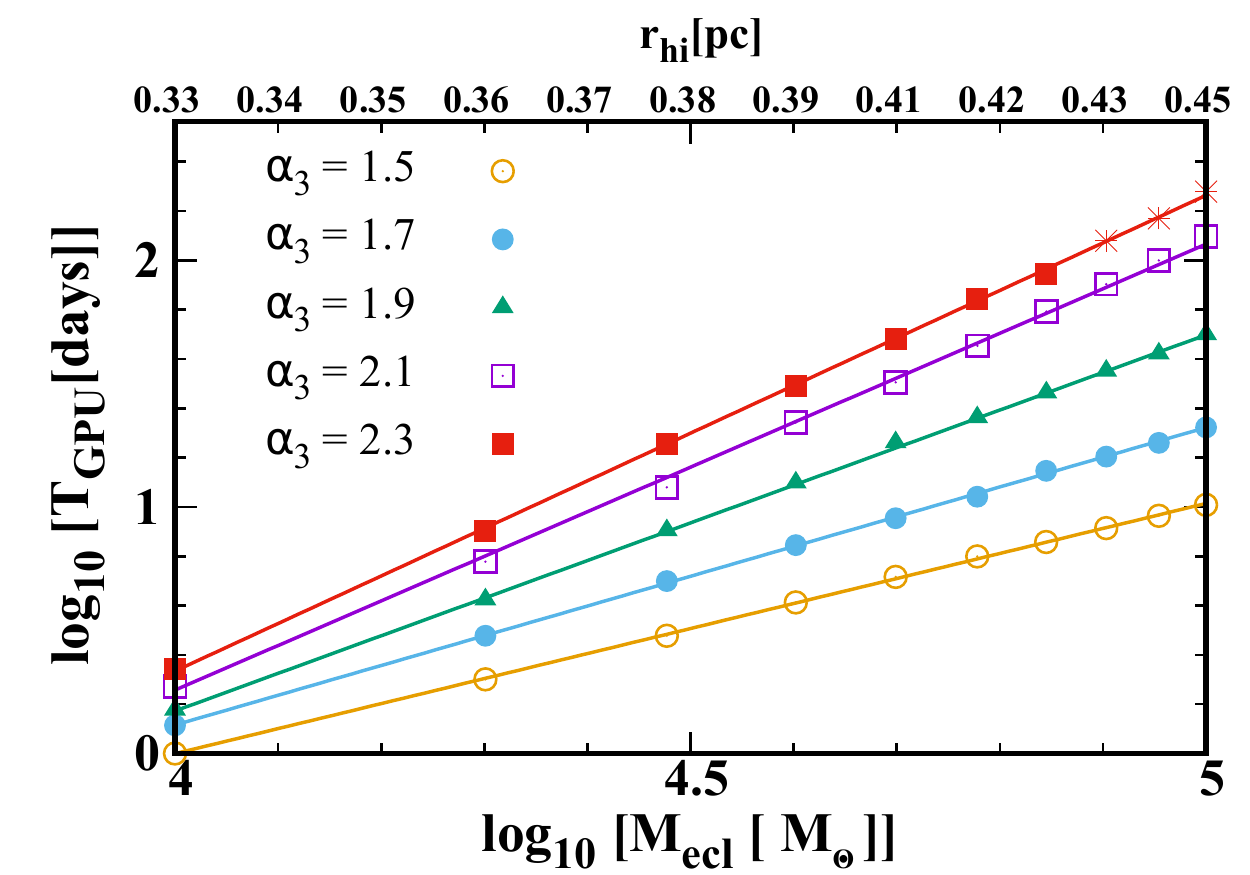}
	\includegraphics[width=85mm]{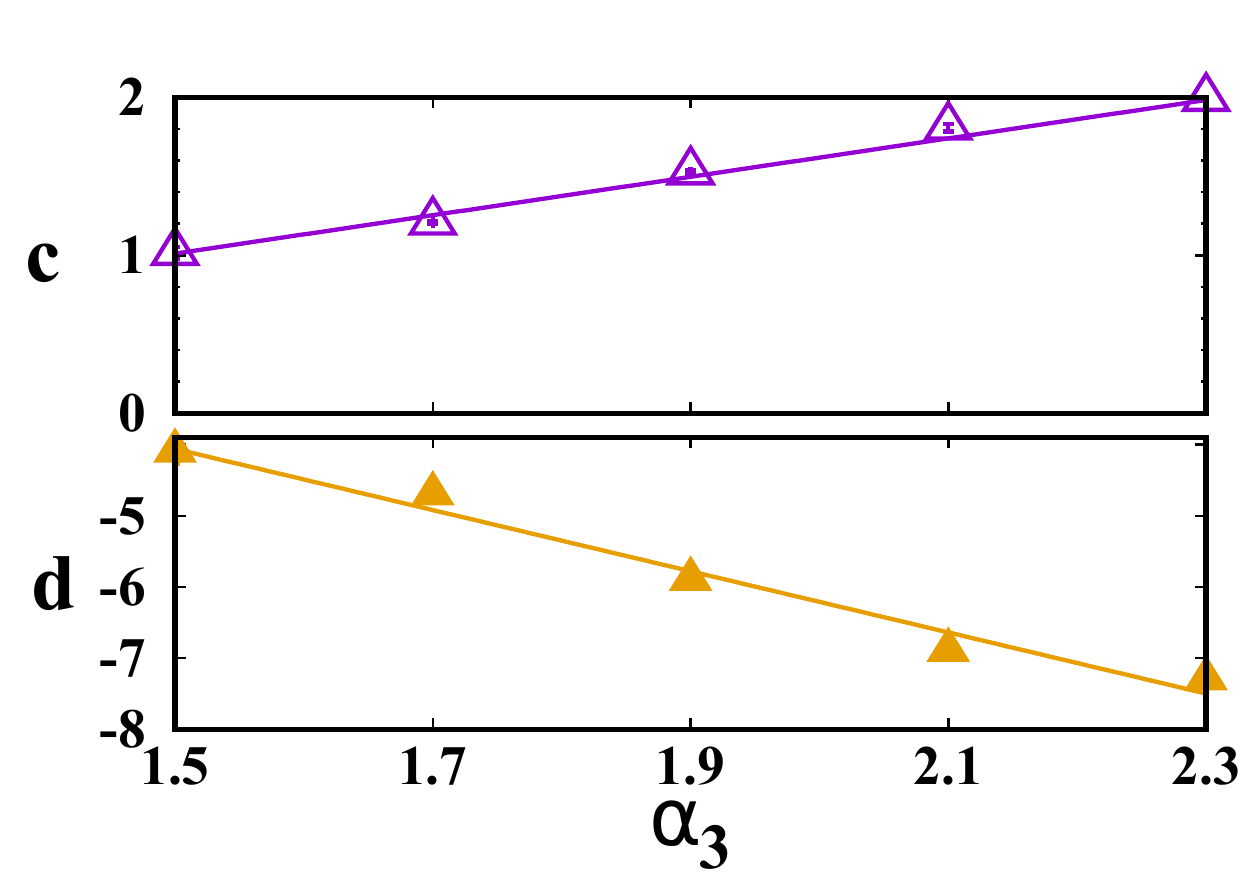}
	\caption{Left panel: GPU run time  as a function of the initial mass of the clusters. The value for the initial half-mass radius \cite{MarksKroupa2012} is given at the top of the panel. Each point indicates the result of one simulation, and the figure shows the required simulation time for each cluster until the dissolution time (Table \ref{tab:appendix-fit}). Solid lines show the linear best-fit (Equation (\ref{equ:appendix-tgpu-m})). The solid lines are the interpolation between the results of our models.  Right panel: The values for the linear-fit parameters of the left panel. The top right panel shows the slope of GPU run time calculated from the linear fit for the results in the left panel as a function of $\alpha_3$. Each point shows one slope for specific $\alpha_3$ and the purple solid line is Equation (\ref{equ:appendix-c}). The bottom right  panel depicts the vertical intercept for the fitting formula in Equation (\ref{equ:appendix-tgpu-m}). The solid line is the best linear fit to these points and is given by Equation (\ref{equ:appendix-d}).  The solid lines in both panels are the linear fits to data  derived from Equation (\ref{equ:appendix-tgpu-m}). }\label{fig:appendix-tgpu-fit}
	\end{figure}

Figure \ref{fig:appendix-tgpu-fit} shows the required time for simulation on a GPU system, $T_{GPU} [days]$, as a function of the initial masses of the embedded clusters, $M_{ecl}$, for different values of  $\alpha_3$. As can be seen, the required time for a simulation on the GPU system  increases when $M_{ecl}$ is  heavier. As a result of the increase in the initial number of stars, the slope of the GPU run time rises in runs with higher $\alpha_3$. The number of heavy stars in clusters with an initially  top-heavy IMF is more than for clusters with a canonical IMF; hence, they will be depleted in the first megayear of the cluster's lifetime.   We can fit the GPU run time by the formula
\begin{equation}
	\log_{10}(T_{GPU})=c(\alpha_3)~\log_{10}(M_{ecl})+d(\alpha_3),
	\label{equ:appendix-tgpu-m}
	\end{equation}
where the best-fit values for each $\alpha_3$ are given in Table \ref{tab:appendix-fit}. The rising lines in Figure \ref{fig:appendix-tgpu-fit} are this fit to the data.
%%
%%%%%%%%%%
%%%%%%
%%

The right panels in Figure \ref{fig:appendix-tgpu-fit} depict the $\alpha_3$ dependence of the coefficients $c$ and $d$ in Equation (\ref{equ:appendix-tgpu-m}) which are the slope of the linear fit and the vertical intercept, respectively. The following functions, $c(\alpha_3)$ and $d(\alpha_3)$, are linear fits to the data points: 
\begin{equation}
	c(\alpha_{3})=(1.2\pm  0.1)~\alpha_{3}-(0.8\pm  0.2),
	\label{equ:appendix-c}
	\end{equation}
%%-0.813018 + 1.21585 *x
\begin{equation}
	d(\alpha_{3})=(-4.3\pm  0.4)~\alpha_{3}+(2.4 \pm 0.7).
	\label{equ:appendix-d}
	\end{equation}
%%2.4 - 4.3 *x
Therefore, one can estimate the required GPU run time for different $\alpha_3$ by extrapolation from our results. For our calculations we used two hardware systems, A and B, where the specifications are summarized in Table \ref{tab:appendix-hardware}.
%% %%%%%%%%%%

%\setcounter{table}{0}
\begin{table}%[]
	\centering
	\caption{GPU run time for $M_{ecl}=$ 100k $M_\odot$ and the Fitting Parameters for Equation (\ref{equ:appendix-tgpu-m}). The first column gives $\alpha_3$ and the GPU run time that is required to simulate a cluster with an initial mass of 100k $M_\odot$ for different $\alpha_3$ is given in the second column. The next two columns denote the fitting parameters for the coefficients $c$ and $d$ in Equation (\ref{equ:appendix-tgpu-m}).  }
	\label{tab:appendix-fit}
	\begin{tabular}{cccc}
		\tablewidth{0pt}
		\hline
		\hline
		$\alpha _ 3$  & $T_{GPU}[days]$ & $c(\alpha_3)$ & $d(\alpha_3)$ \\
		\hline
		\decimals
		1.5  & 10&1.01  $\pm$ 0.04& -4.11  $\pm$ 0.20  \\%1.05*x-1.8
		1.7  & 21&1.21  $\pm$ 0.01 & -4.71  $\pm$ 0.04 \\%1.0213*x-1.2144
		1.9  & 50& 1.50 $\pm$ 0.01& -5.90 $\pm$ 0.05  \\ %0.91*x-0.306
		2.1  & 125& 1.80 $\pm$ 0.02 & -6.90 $\pm$ 0.11  \\%0.85935233*x+0.094341016
		2.3  &190& 1.90 $\pm$ 0.01& -7.30 $\pm$ 0.06  \\ %0.8236923*x+0.319572730
		\hline
		\end{tabular}
	\end{table}

%% %%%%%%%%%%
\begin{table}  %[ht]
	\centering
	\caption{ Hardware and Software Configurations.}
	\label{tab:appendix-hardware}
	\begin{tabular}{lcc}
		\hline
		\hline
		& System A & System B \\
		\hline
		CPU &Core i7-6900(3.2 GHz)& Core i7-3820(3.6 GHz)\\
		GPU &Geforce GTX 1080&Geforce GTX 690\\
		Motherboard & ASUS X99-PRO&  ASUS X79\\
		Memory & 16 GB & 8 GB  \\
		OS & CentOS 7& CentOS 6  \\
\hline
		\end{tabular}
	\end{table}
%%%%%%

\end{document}